\documentclass[usenatbib]{mnras}
\usepackage[utf8]{inputenc}
\usepackage{amsmath}
\usepackage{amssymb} 
\usepackage{graphicx}
\usepackage{caption}
\usepackage{subcaption}
\usepackage{ulem}
\PassOptionsToPackage{hidelinks}{hyperref}

\title[FRBs from non-resonant Alfv\'en waves]{Fast Radio Bursts from non-resonant Alfv\'en waves and synchrotron maser emission in the magnetar wind}
	
\author[K. Long \& A. Pe'er]{
Killian Long,$^{1}$\thanks{E-mail: killian.long@umail.ucc.ie (KL)}
Asaf Pe'er,$^{2}$
\\
$^{1}$Department of Physics, University College Cork, Cork, Ireland\\
$^{2}$Bar-Ilan University, Ramat-Gan 5290002, Israel
}

\date{Accepted XXX. Received YYY; in original form ZZZ}

\pubyear{\the\year{}}
\begin{document}

	
\maketitle
	\begin{abstract}
	Non-resonant interactions between Alfv\'{e}n waves and a relativistic plasma result in the formation of the population inversions necessary for synchrotron maser emission (SME) across a wide range of magnetisations and temperatures. We calculate the peak frequencies of the SME resulting from this interaction and show that the characteristic frequencies and energetics of fast radio bursts (FRBs) can be produced in the relativistic wind of a magnetar using this mechanism. Wind Lorentz factors of $\gamma_w\gtrsim310$ are shown to be necessary to explain observed FRBs. Emission is possible at temperatures of $\theta = k_bT/mc^2\lesssim 0.02$. We further examine the periods and magnetic fields of the central magnetar and demonstrate that the optimal values of these properties align with the observed magnetar population provided that the magnetosphere is disturbed by the flaring activity. These results allow the properties of the environment such as temperature and magnetisation to be probed from the observed FRB frequency and luminosity.
	\end{abstract}
    

 \section{Introduction}
 \label{sec:intro}

Fast Radio Bursts (FRBs) are extremely bright transients of millisecond duration in the GHz frequency range and brightness temperature reaching $T_b \sim 10^{36}\text{ K}$ \cite[e.g.][]{2016MPLA...3130013K}. 
The nature of the physical processes that produce FRBs is to date still debatable, with many theoretical ideas \citep[for a summary of models, see, e.g.,][]{2019PhR...821....1P}. To reach such extreme brightness temperatures, a coherent emission mechanism is required. One such mechanism is synchrotron maser emission (SME), which has been invoked to explain FRBs \citep[e.g.][]{Lyubarsky2014,2018ApJ...864L..12L,Metzger2019} as well as other coherent signals including Auroral Kilometric Radiation (AKR) in the Earth's magnetosphere \citep{1979ApJ...230..621W, Wu1985}, Jovian Decametric Radiation \cite[e.g.][]{1982AuJPh..35..447H}, and blazars \citep{2005ApJ...625...51B}. SME requires a background magnetic field and an inverse population of electrons such that the energy $E$ of the particle distribution grows faster than $E^2$ \citep{2002ApJ...574..861S} or satisfies $\frac{\partial F}{\partial p_{\perp}}>0$ \citep{ Wu1985}. Here $F(p_{\perp},p_{\parallel})$ is the particle distribution function and $p_{\perp} (p_{\parallel})$ is the momentum perpendicular (parallel) to the background magnetic field $\mathbf{B}$.

 It has been shown that such a distribution can be achieved through the non-resonant interaction of Alfv\'en waves with both non-relativistic \citep{Zhao2013, Wu2014} and relativistic \citep{2023PhRvD.107l1301L, 2025MNRAS.538.1029L} plasmas. In these works, it was shown that pitch-angle scattering produces a crescent-shaped population inversion over a wide range of temperatures $\theta = \frac{k_BT}{m_ec^2} \leq 3$ in an electron-positron plasma. Here, $k_B$ is the Boltzmann constant, $T$ is the temperature, $m_e$ is the electron mass, and $c$ is the speed of light in vacuum. The population inversion forms more efficiently at higher magnetisations of $\sigma = \frac{B'^2}{4\pi w m_ec^2}\gtrsim1$ for temperatures greater than $\theta \sim 10^{-2}$ \citep{2025MNRAS.538.1029L}, where $B'$ denotes the magnetic field in the plasma rest frame and $w$ is the proper specific enthalpy density. 
 
While these works lay the physical ground for associating FRBs with non-resonant interaction between Alfv\'en waves and plasma, a complete physical scenario that could link FRBs with known objects such as magnetars is still lacking. In this work we propose such a model and examine the conditions needed for producing the detected FRB signal. As we show below, FRBs can be produced through non-resonant interactions between Alfv\'en waves and the relativistic magnetar wind. These interactions produce a crescent-shaped population inversion capable of supporting SME.
 
This paper is organized as follows. In Section \ref{sec:model} we describe the FRB model, detailing the parameters of the magnetar wind and establishing the parameter space where SME can occur. These results are used to constrain the magnetar and wind properties in Section \ref{Sec:constraints}. Section \ref{sec:frb} describes how the model can explain FRB 20200428, the only FRB with a confirmed magnetar origin. Section \ref{sec:dis} contains the discussion, and finally the results are summarised in the conclusion in Section \ref{sec:conclusions}.

\section{Model}
\label{sec:model}
Magnetars are the most popular progenitors of FRBs due to their large reserves of magnetic and rotational energy and compact scales \citep[e.g.][]{Zhang2020}. This link has been strengthened by FRB 20200428, which has been observed to originate from the galactic magnetar SGR 1935+2154 \citep{2020Natur.587...54C, 2020Natur.587...59B}. The surface magnetic field $B_*$ of magnetars is extremely high, with typical values of $B_*\sim10^{14-15}\text{ G}$ \citep[e.g.][]{2017ARA&A..55..261K}. The magnetar is surrounded by a highly magnetised magnetosphere in which $B\propto R^{-3}$. This closed magnetosphere extends to the light cylinder radius $R_{LC} = cP/(2\pi) = 4.8\times10^9 P \text{ cm}$, where $P$ is the rotational period of the magnetar \citep{1969ApJ...157..869G}. The magnetic field at this radius is therefore $B_{LC} = B_*R_*^3 R_{LC}^{-3} = 9\times10^3\mu_{33}P^{-3}\text{ G}$. Here, $R_*$ is the magnetar radius, $\mu\sim B_*R_*^{3}$ is the magnetic moment of the neutron star, and we have used the convention $Q = 10^x Q_x$ in cgs units. We have assumed typical values of $B_*=10^{15}\text{ G}$ and $R_{*} = 10^{6}\text{ cm}$. Beyond the light cylinder, the magnetic field lines open and electron-positron pairs are ejected from the magnetosphere to form a magnetar wind. The magnetic field in the wind is dominated by the azimuthal component $B_\phi\propto R^{-1}$ \citep[e.g.][]{2017SSRv..207..111C}, so that the field at a radius $R>R_{LC}$ is given by $B(R) = 44 \mu_{33}P^{-2}R_{12}^{-1}\text{ G}$. This wind is the region where SME is instigated either by the non-resonant interaction of Alfv\'en waves with the wind plasma \citep{2025MNRAS.538.1029L} or by the presence of plasma ejected in a flaring event, as we show in detail below. This SME can result in the production of an FRB.

The wind is launched at the light cylinder, where the magnetisation is $\sigma_{LC}  = B_{LC}^2 /(4 \pi n m_e c^2) = B_{LC}^2 R_{LC}^2/(\dot{N} m_ec) $ in the case of a cold plasma, with $n$ denoting the total particle density and $\dot{N}$ denoting the particle flux. This flux can be estimated as $\dot{N} \sim 3\times10^{39}\mathcal{M}_3\mu_{33}^{2/3}P^{-1}\text{ s}^{-1}$ \citep{2020ApJ...896..142B}, resulting in a magnetisation of $\sigma_{LC} \sim 3 \times10^{4}\mu_{33}^{4/3}\mathcal{M}_3^{-1}P^{-3}$. Here, $\mathcal{M}$ is the pair multiplicity. The wind is initially accelerated linearly with radius by fast magnetosonic waves until the flow reaches the fast magnetosonic surface, at which its bulk Lorentz factor is $\gamma_{fms}\sim\sigma_{LC}^{1/3}$. Acceleration beyond this point is slower, with the wind reaching an asymptotic Lorentz factor of $\gamma_w\sim 3\sigma_{LC}^{1/3}$ \citep{1969ApJ...158..727M, 1998MNRAS.299..341B}. 

The wind far from the equator is expected to be relatively cold ($\Delta \gamma_w/\gamma_w<<1$) \citep{2020ApJ...896..142B}. Closer to the equator reconnection in the striped wind may result in more efficient acceleration and heating, with Lorentz factors of up to $\gamma_w\sim \sigma_{LC}$ possibly being achieved \citep{2001ApJ...547..437L}. Although the details of the wind parameters are only estimates due to the uncertainties in the efficiencies of the various dissipation processes and in the value of the pair multiplicity, magnetar winds of this type have magnetisations of $\sigma\geq1$ and are highly relativistic ($\gamma_w>>1$) for large radii $R>>R_{LC}$. It is also likely that an enhanced pre-flare wind \citep{2020ApJ...896..142B} or the impact of the starquake \citep[e.g.][]{2023MNRAS.524.6024S} will perturb the magnetic field configuration in the magnetosphere, which will result in enhanced magnetic fields in the wind region itself.

In order for a population inversion to form in such a wind through non-resonant interactions, Alfv\'en waves from the central magnetar must propagate through the magnetosphere to the emission region. Results from \cite{2025MNRAS.538.1029L} show that the ratio of the energy density of the Alfv\'en waves to the background magnetic field, $\eta = B_W^2/B^2$, must have a magnitude of $\eta \gtrsim 0.1$ at the emission site for a significant population inversion to form. Starquakes, which occur due to shear oscillations of the magnetar crust, can launch the required Alfv\'en waves into the inner magnetosphere \citep{1989ApJ...343..839B, 1996ApJ...473..322T}. The typical energy released by a starquake is estimated to be on the order of $E_Q\sim10^{42}\text{ erg}$ \citep[e.g.][]{2025arXiv250812567Q}, although larger energies of $E_Q\gtrsim10^{44}\text{ erg}$ may also be attainable \citep[e.g.][]{2020ApJ...900L..21Y}. Only a fraction of this energy is emitted in the form of Alfv\'en waves, with transmission coefficients of $10^{-2}-10^{-1}$ \citep[e.g.][]{2020ApJ...897..173B, 2025arXiv250812567Q}, resulting in a small initial relative energy density of $\eta<<1$. However, 
as the magnetic field of the Alfv\'en wave decreases more slowly than the background field in the magnetosphere, the relative strength of the turbulence grows with radius as $\eta(R) = \eta_*(R/R_*)^{3/2}$ \citep[e.g.][]{2020ApJ...900L..21Y}. Here, $\eta_*$ is the turbulence level at the magnetar surface. To reach values of $\eta>0.1$ at the light cylinder therefore requires only small initial values of $\eta_*>3\times10^{-7}P^{3/2}R_{*,6}^{3/2}$, suggesting that the turbulence levels needed to form a population inversion are readily achievable in this scenario.

The natural time-scale for the growth of the Alfv\'en waves from the starquake is some fraction $\chi$ of the magnetar radius time-scale, $t_A\sim\chi R_*/c=3.3\times10^{-5}\chi R_{*,6}\text{ s}$. In other words, the turbulence level $\eta$ reaches its asymptotic value at $t\sim t_A$. This is also the time-scale for the formation of the population inversion, as it has been shown for both the non-relativistic \citep{Wu2007,Yoon2009} and relativistic \citep{2025MNRAS.538.1029L} cases that the asymptotic distribution depends only on the turbulence level for a given set of parameters. This scale is much smaller than the typical length scales of the magnetar wind, in which $B\propto R^{-1}$ for $R>>R_{LC}$.

\subsection{Parameter space for SME}
\label{sec:maser}
\subsubsection{Growth rate calculation - general considerations}

\begin{figure}
	\centering 
	\includegraphics[width = 1\columnwidth]{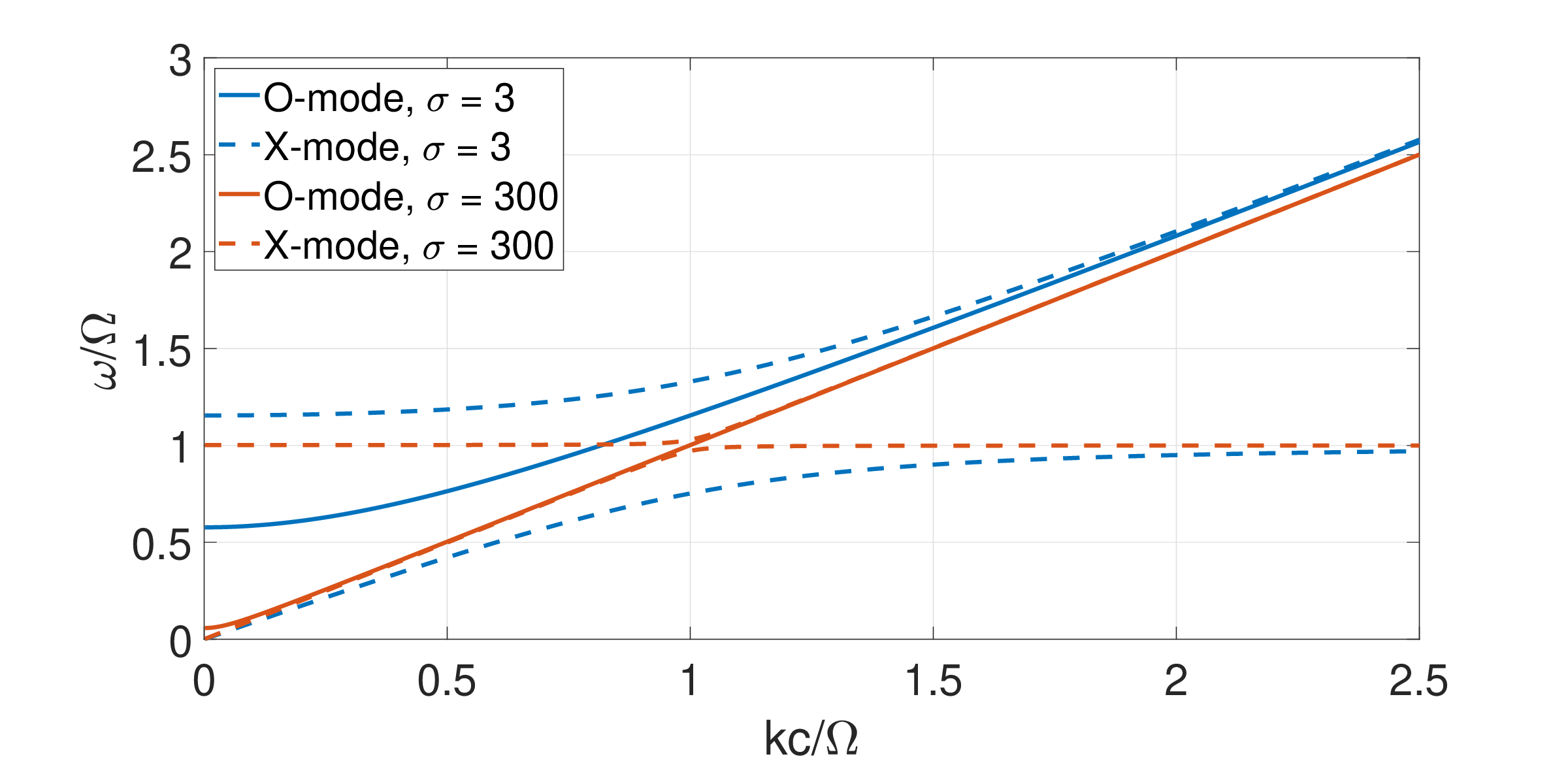}
	\caption{The dispersion relation for perpendicular propagation in a cold electron-positron plasma with $\sigma = 3$ is shown in blue, and $\sigma=300$ in orange. The solid lines show the ordinary mode dispersion relation, while the extraordinary modes are shown by dashed lines. The ordinary mode has a cutoff at $\omega = \omega_p$, which has a value of $\omega_p =0.58 \Omega$ for $\sigma = 3$ and $\omega_p = 5.8\times10^{-2}\Omega$ for $\sigma=300$. The extraordinary mode has a cutoff at $\omega = \left(\omega_p^2+\Omega^2\right)^{1/2}$ and a resonance at $\omega = \Omega$. The dispersion relation becomes more complicated for oblique propagation due to the introduction of further resonances.}
	\label{fig:ox}
\end{figure}

 In the presence of a population inversion, electromagnetic waves can resonate with these particles and become amplified, resulting in maser emission. The modes in question are typically either the ordinary or extraordinary electromagnetic wave modes. An example of a dispersion relation for these modes in a magnetised plasma is shown in Fig. \ref{fig:ox} for propagation perpendicular to the magnetic field. The ordinary mode has a cutoff at $\omega = \omega_p$, while the extraordinary mode has a cutoff at $\omega = \left(\omega_p^2+\Omega^2\right)^{1/2}$ and a resonance at $\omega = \Omega$ \citep[e.g. ][]{1993PhRvE..47..604I}, where $\omega_p=\sqrt{4\pi n e^2/m_e}$ is the plasma frequency and $\Omega = eB/m_ec$ is the cyclotron frequency. The extraordinary mode has upper and lower branches, which we refer to as the fast and slow branches respectively. The wave frequency has real and imaginary components such that $\omega = \omega_r+i\omega_i$. SME occurs when the imaginary part of the frequency, the growth rate $\omega_i$, is positive. As masing is a resonant process, the non-resonant Alfv\'en waves will not be amplified.  We analyse the conditions required for successful maser emission below.

 To simplify the calculation of the growth rate, we assume that the emitted waves propagate through a thermal plasma (denoted by the subscript $0$), and that the inverse population of energetic electrons (subscript $e$) only affects the growth rate of the waves. We do not expect the peak frequency to be significantly altered by this assumption, although the precise values of the growth rate will change. The growth rate is calculated using the expression, \citep[see e.g.][]{Wu1985, 1992wapl.book.....S}
 
  \begin{equation}
      \omega_i = -\frac{ \text{Im}(\Lambda)}{\frac{\partial}{\partial \omega}(\text{Re}(\Lambda_0))}.
  \end{equation}

\begin{figure}
	\centering 
       \captionsetup{width=\columnwidth}
	\includegraphics[width = \columnwidth]{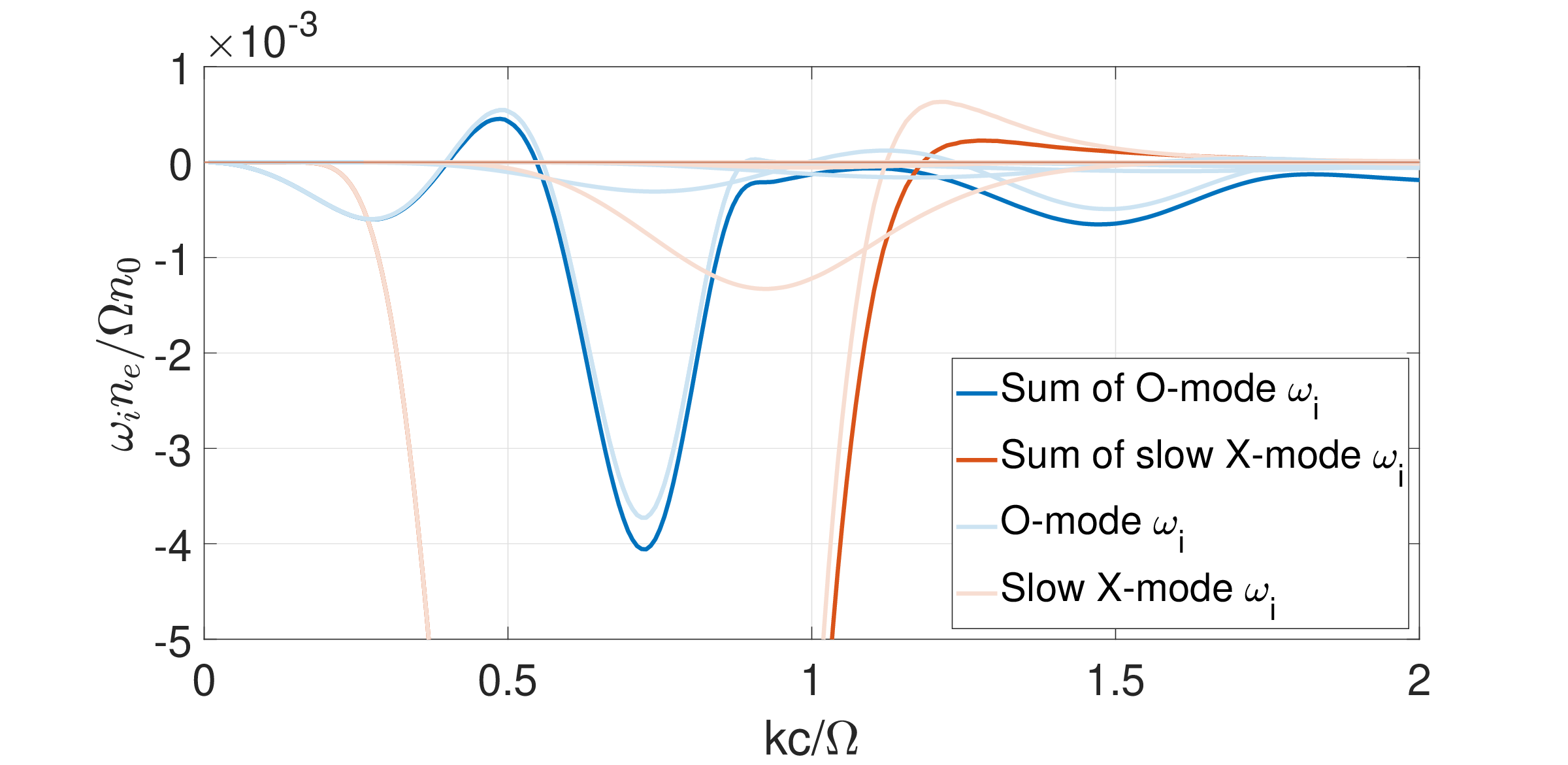}
	\caption{The growth rate $\omega_i$ normalised to the cyclotron frequency and density ratio for the first 10 harmonics of the O- and slow X-modes, with the majority of the contribution coming from the first 3 harmonics in both cases. The results are for a crescent shaped distribution with initial temperature $\theta=10^{-2}$, $\sigma = 10$ and turbulence level $\eta=1$. The individual ordinary mode harmonics are shown as faint blue lines, with the sum over all harmonics shown as the solid blue line. The same is the case for the extraordinary mode harmonics which are shown in orange. The fastest growing wavenumber for the O-mode occurs at $k\sim0.48\Omega/c$, while for the slow X-mode it is found at the higher wavenumber of $k\sim1.27\Omega/c$. The range of wavenumbers where the growth rate is positive is broadened due to the temperature of the distribution in comparison to a cold scenario. The overlap of harmonics due to this broadening acts to reduce the growth rate, especially for higher harmonics.}
	\label{fig:modes}
\end{figure}

  Here $\Lambda = \text{det } \mathbf{\Lambda}$, and $\Lambda_0$ refers to the thermal plasma, while $\Lambda$ contains the contributions from both the thermal and energetic particles. The components of the dispersion relation $\mathbf{\Lambda}$ are given by 

  \begin{equation}
      \Lambda_{ij} = N_r^2\left(\frac{k_ik_j}{k^2}-\delta_{ij}\right)+\varepsilon_{ij},
  \end{equation}
  with $N_r$ denoting the refractive index, $k$ the wavenumber, $\delta_{ij}$ the Kronecker delta function and $\varepsilon$ the dielectric tensor. The full details of this calculation and the terms involved are provided in Appendix A. In general, positive growth will be achieved for a range of wavenumbers, as can be seen, for example, in Fig. \ref{fig:modes}, which shows the first 10 harmonics of both the ordinary and slow extraordinary modes for $\theta=10^{-2}$ and $\sigma=10$. For this set of parameters there is no positive growth for the fast extraordinary mode. As the signal grows exponentially, the fastest growing mode will dominate the process. We designate the real part of the frequency of this mode as the emission frequency $\omega_m$ and the growth rate of the mode as $\Gamma_i$.

  We calculate $\Gamma_i$ as well as the peak emission frequency $\omega_m$ for temperatures in the range $10^{-3} \leq \theta \leq 3$ and magnetisations $\sigma > 1$. The peak emission frequency is typically $\omega_{m}' \sim l \Omega'$, where $l$ is a temperature and magnetisation dependent numerical factor whose value is determined by the dominant mode of emission \footnote{Due to the overlap of modes and relativistic effects $l$ does not have integer values.} and $\Omega'$ is the cyclotron frequency as measured in the plasma rest frame, given by $\Omega' = eB'/m_ec$. Throughout this work, primed quantities are measured in the plasma rest frame, while unprimed quantities are measured in the observer's frame.

\begin{figure*}
	\centering 
       \captionsetup{width=2\columnwidth}
{\phantomsubcaption\label{fig:disa}%
\phantomsubcaption\label{fig:disb}%
    }

	\includegraphics[width = 2\columnwidth]{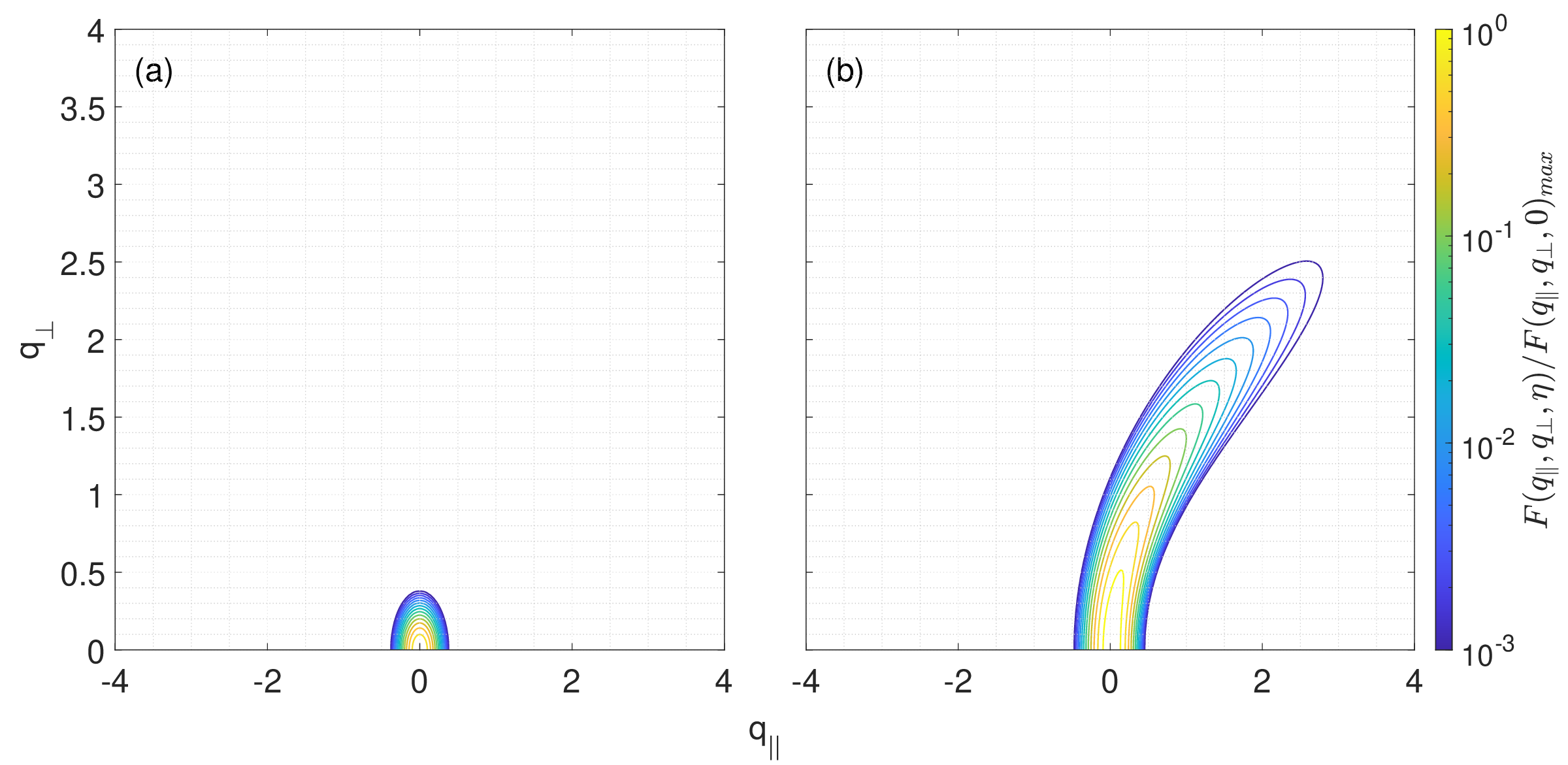}
	\caption{An example distribution $F(q_\parallel,q_\perp,\eta)$ for $\theta = 10^{-2}$ and $\sigma = 300$. Panel (a) shows initial distribution at $\eta=0$ (where no non-resonant interaction has yet occurred) and panel (b) shows the distribution at a turbulence level of $\eta=1$. Both plots are normalized to the maximum value of $F(q_\parallel,q_\perp,0)$ for presentation purposes. In panel (b), the formation of a crescent shape is clear, with particles gaining energy in the region of positive parallel and perpendicular momentum. Note that the increase in the width of the distribution in the parallel direction is much lower than the increase in the perpendicular direction. }
	\label{fig:dis}
\end{figure*}

 \subsubsection{Formation of the population inversion and resulting growth rates}
 \label{sec:grcalc}
In our model, the Alfv\'en waves interact with the particles in the wind, which are assumed to have an initial Maxwell-J\"uttner distribution with temperature $\theta$. This thermal distribution does not result in any maser emission as it does not contain a population inversion. The non-resonant interaction deforms the initial distribution into a crescent-shaped distribution with enhanced kinetic energy and a population inversion for $p_\parallel>0$ as shown in \cite{2025MNRAS.538.1029L}. We show the initial distribution at $\eta=0$, i.e. before any Alfv\'enic turbulence, in Fig. \ref{fig:disa} for $\theta=10^{-2}$. As the Alfv\'en wave turbulence grows this initial distribution is deformed, resulting in the particle distribution shown in Fig. \ref{fig:disb} once the turbulence reaches an asymptotic value of $\eta=1$. As described above, this occurs on a time-scale of $t\sim3.3\times10^{-5}\chi R_{*,6}\text{ s}$. The quantity $q$ shown in Fig. \ref{fig:dis} is the normalised momentum, $q=p/mc$. 

\begin{figure}
	\centering 
       \captionsetup{width=\columnwidth}
{\phantomsubcaption\label{fig:growth_tempa}%
\phantomsubcaption\label{fig:growth_tempb}%
    }

	\includegraphics[width = \columnwidth]{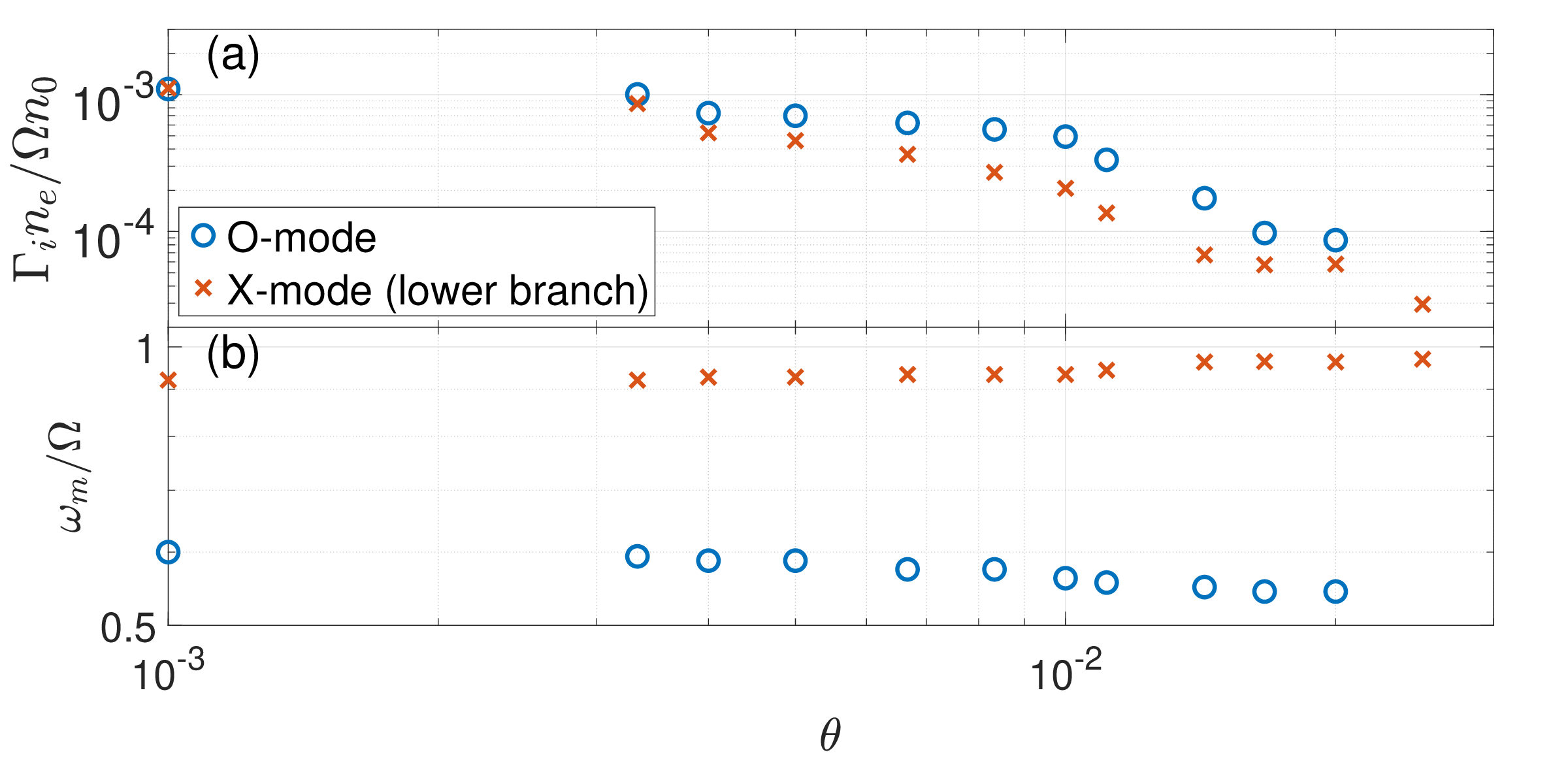}
	\caption{Growth rate and emission frequency as a function of temperature: Panel (a) shows the growth rate $\Gamma_i$ normalised to the cyclotron frequency and density ratio for $\sigma=10$ and $\eta=1$, while panel (b) shows the peak emission frequency $\omega_m$, also normalised to the cyclotron frequency. The ordinary mode is shown by blue 'o's and the slow extraordinary mode (lower branch) by orange 'x's. The decrease in growth rate with increasing temperature is clearly visible. For temperatures of $\theta>0.02$ no maser emission occurs for the O-mode, while the emission in the X-mode continues to slightly higher temperatures of $\theta \sim 0.025$. Above these temperatures no growth was found. At this magnetisation the O-mode is dominant at temperatures of $\theta\gtrsim10^{-3}$, at which point the growth rates become comparable. In all cases the maximum growth rate is found at a propagation angle of $\pi/2$.}
	\label{fig:growth_temp}
\end{figure}

The normalised growth rates for the crescent distribution at a turbulence level of $\eta=1$ are shown for varying temperatures and a magnetisation of $\sigma = 10$ in Fig. \ref{fig:growth_tempa}, with the emission frequencies also shown in Fig. \ref{fig:growth_tempb}.
 These growth rate calculations show that the maser emission is only possible for initial temperatures of $\theta\lesssim0.02$ in this model, as can be seen in Fig. \ref{fig:growth_tempa}. At higher temperatures, no positive growth rate was obtained for any mode. This temperature restriction is compatible with typical descriptions of the magnetar wind, which as described above is modelled as a cold pair plasma moving relativistically \citep{2020ApJ...896..142B}. In order for a given wave mode to have a positive growth rate, the mode must primarily resonate with particles in the population inversion, i.e. those located in the region of momentum space where $\partial F/\partial p_\perp >0$. If the wave mode is also in resonance with particles where $\partial F/\partial p_\perp <0$, absorption rather than growth will occur, resulting in damping of the waves. At higher temperatures, the width of the particle distribution results in the resonance ellipses for a given wave mode interacting with regions of $\partial F/\partial p_\perp <0$ that outweigh any growth from the population inversion. This means that even though there is still an inverse population of energetic particles, the contribution to wave growth from these particles is insufficient to overcome the damping effects of the rest of the distribution. The details of the resonance are discussed in further detail in Appendix A. 

\begin{figure}
	\centering 
       \captionsetup{width=\columnwidth}
{\phantomsubcaption\label{fig:growth_maga}%
\phantomsubcaption\label{fig:growth_magb}%
    }

	\includegraphics[width = \columnwidth]{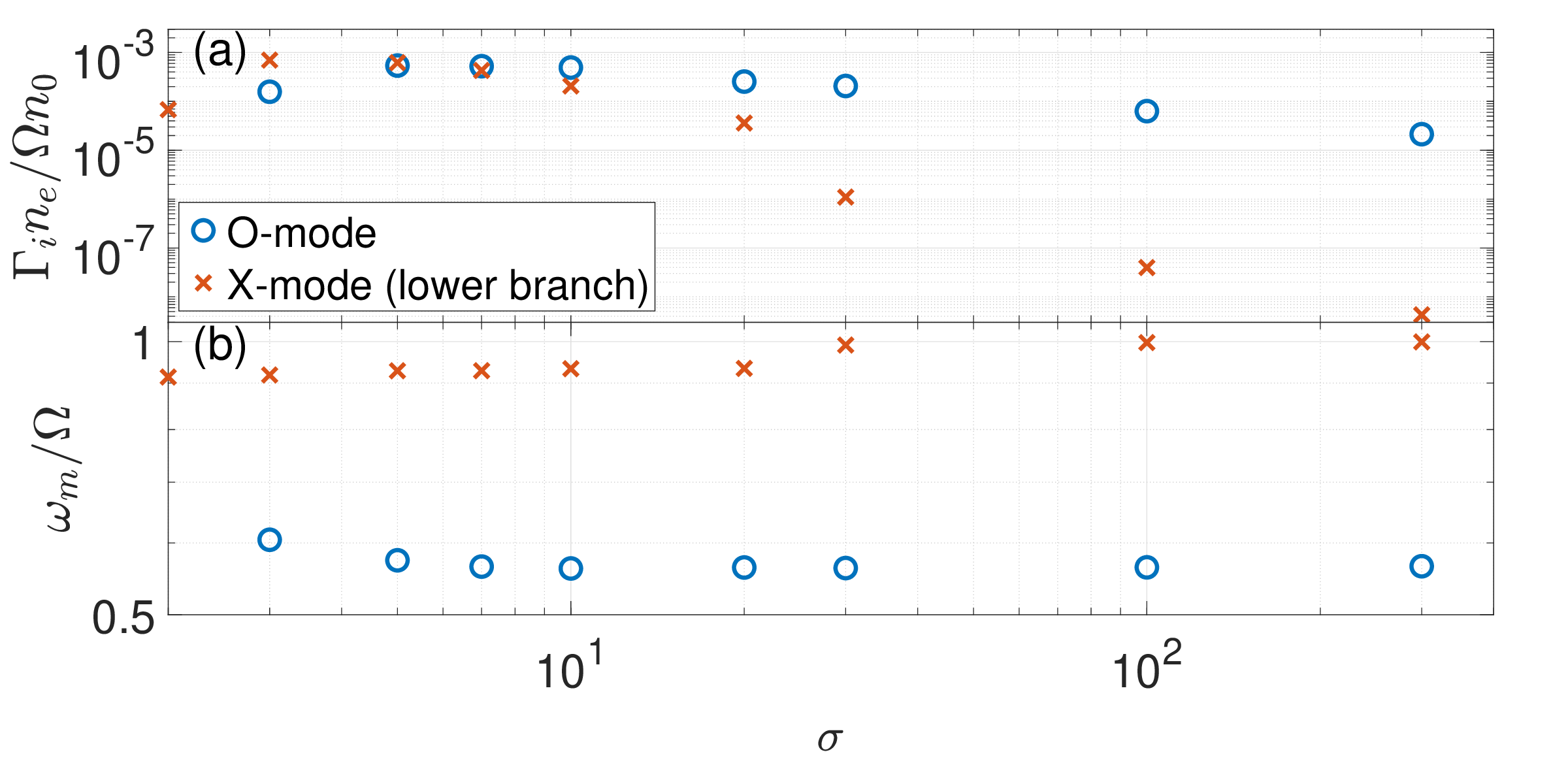}
	\caption{Growth rate and emission frequency as a function of magnetisation: Panel (a) shows the growth rate $\Gamma_i$ normalised to the cyclotron frequency and density ratio for $\theta=10^{-2}$ and $\eta=1$, while panel (b) shows the peak emission frequency $\omega_m$, also normalised to the cyclotron frequency. The ordinary mode is shown by blue 'o's and the slow extraordinary mode (lower branch) by orange 'x's. The decrease in growth rate with increasing magnetisation due to the reduced number density is clearly visible. The O-mode is dominant at magnetisations of $\sigma\gtrsim5$ for this temperature, with no O-mode growth present for $\sigma<2$. The crossover value increases with decreasing initial temperatures. In all cases the maximum growth rate is found at a propagation angle of $\pi/2$.}
	\label{fig:growth_mag}
\end{figure}

The dominant mode of emission depends on both the magnetisation and temperature. We show the normalised growth rates and emission frequencies as a function of $\sigma$ for $\theta=10^{-2}$ in Fig. \ref{fig:growth_maga} and \ref{fig:growth_magb}. At lower magnetisations of $\sigma\lesssim3-30$ the lower branch of the extraordinary mode (the 'slow' mode) dominates, as shown in Fig. \ref{fig:growth_maga}. The slow extraordinary mode is a trapped mode, and thus cannot be observed unless mode conversion occurs, as has been proposed for various phenomena such as solar coronal emission \citep[e.g. ][]{2012ApJ...746...68K}.
At higher magnetisations the ordinary mode dominates, with the crossover magnetisation increasing as the temperature decreases. Emission from the upper branch of the extraordinary mode is never favoured for the parameters examined due to its frequency always being higher than the cyclotron frequency. Particles in the population inversion have relativistic momenta and hence lower gyration frequencies, which results in less efficient interactions with upper branch of the X-mode.

As can be seen in Fig. \ref{fig:modes}, the slow extraordinary mode has positive growth in regions of $k\gtrsim\Omega/c$, corresponding to emission frequencies of $\omega_m > 0.9\Omega$ (see the dispersion relation presented in Fig. \ref{fig:ox} and the results for $\omega_m$ in Fig. \ref{fig:growth_tempb} and \ref{fig:growth_magb}).
In contrast, the O-mode favours growth at lower frequencies of $\omega_m\sim0.55-0.6\Omega$, which corresponds to resonance ellipses of greater radii in momentum space (see the factors of $\frac{n\Omega}{\omega_r}$ in Eq. \ref{eq:resel}). This implies that O-mode growth is most efficient when the waves resonate with particles of higher momenta. As the ordinary mode has a lower cutoff at $\omega/\Omega=\omega_p/\Omega\sim\sigma^{-1/2}$, the growth rate of the O-mode is reduced at lower magnetisations of $\sigma\lesssim5$, resulting in the mode becoming less dominant in this region of the parameter space, as can be seen in Fig. \ref{fig:growth_maga}. For both the O and X-modes, the normalised growth rate ($\Gamma_i/\Omega$) decreases with increasing magnetisation for $\sigma\gtrsim 10$ due to the decreasing number density, as can be seen in Fig. \ref{fig:growth_maga}. The growth rate also decreases as the temperature increases due to the increased width of the distribution (and resulting overlap of modes), as well as the lower gradients. In all cases, modes propagating perpendicularly to the background magnetic field show the highest growth rates, as seen in similar non-relativistic scenarios \citep{Zhao2013}.

 For perpendicular propagation, the electric field of an ordinary mode wave is parallel to $\mathbf{B}$. As such, its behaviour only depends on the parallel element of the dispersion relation (i.e. the $Q_{zz}$ terms in $\Lambda_{zz}$ in Eq. \ref{eq:lambda}). This term is proportional to $p_\parallel^2$. As a result, the O-mode primarily extracts energy from the 'arms' of the crescent distribution because of the larger parallel momentum in this region, leading to lower emission frequencies due to relativistic effects. This dependence also leads to the reduction of growth rates at lower magnetisations, as the increasing value of the lower frequency cutoff prevents resonance between particles with large $p_\parallel$ values and O-mode waves. Furthermore, as the magnitude of this term depends on the spread of the distribution in the parallel direction, it becomes relatively weaker at low temperatures. As can be seen in Fig. \ref{fig:dis}, the width of the population distribution does not change substantially in the parallel direction due to the non-resonant interaction.
 
 In contrast, the perpendicular momentum increases dramatically. The perpendicular momentum of the particles at $\eta=1$ is dominated by the energy gained from the Alfv\'en waves rather than the initial thermal energy. As the X-mode waves have an electric field perpendicular to $\mathbf{B}$, they depend only on the $\Lambda_{yy}$ term in Eq. \ref{eq:aplambda}\footnote{The convention $k_y=0$ has been used throughout Appendix A, hence why the X-mode wave only depends on $\Lambda_{yy}$ and not $\Lambda_{xx}$.}. As this term is proportional to $p_\perp^2$, this results in a relative strengthening of the X-mode relative to the O-mode at lower temperatures. This can be seen in Fig. \ref{fig:growth_maga}, where the growth rates become comparable at $\theta\sim10^{-3}$. The growth rates of the slow X-mode are also not as strongly reduced at low magnetisations compared to those of the O-mode. The decreasing magnetisation does not change the frequency range allowed for the slow extraordinary mode. This mode also always favours resonance ellipses with relatively smaller radii in momentum space. As this mode is proportional to $p_\perp^2$, the regions of negative $\partial F/\partial p_\perp$ at larger momenta result in enhanced damping in comparison to the O-mode, leading to smaller resonance ellipses being preferred.

\section{Constraints}
\label{Sec:constraints}
To examine the emission region, we combine the constraints on the allowed parameter space for SME determined above with three further constraints derived from observed FRB parameters and physical considerations. These are (i) the observed FRB frequency; (ii) the FRB time-scale; and (iii) the requirement that the emission occurs in a region smaller than the coherent length scale of the magnetic field. This is necessary to ensure that the background field and emission frequency do not change substantially across the masing region.
\subsection{Frequency}
\label{Sec:freq}

 FRBs are primarily detected in the GHz band, with individual signals observed from as low as 110 MHz \citep{2019ARep...63...39F,2021ApJ...911L...3P,2021Natur.596..505P} up to 8 GHz \citep{2018ApJ...863....2G}. These observations can be used to constrain the required magnetic field for SME and hence the radius from the magnetar. As shown in Section \ref{sec:maser}, the peak frequency for SME is $\omega_m'=l\Omega'$ in the wind frame. Presuming the relativistic wind and emission are oriented towards the observer, this corresponds to a required magnetic field of 
 \begin{equation}
 B = \frac{\pi\nu}{bl}\frac{m_ec}{e} \sim 179 b^{-1}l^{-1}\nu_9\text{ G},
 \end{equation}
 for GHz frequencies. The term
 \begin{equation}
b = \gamma_wB'/B = \sqrt{\frac{\gamma_w^2\xi^2+1}{\xi^2+1}}\geq 1,
 \end{equation} 
 is a factor that accounts for the orientation of the magnetic field, where $\xi=B_r/B_\phi$. In the far wind ($R>>R_{LC}$), the magnetic field is perpendicular to the bulk velocity of the wind and $b=1$ as $B_\perp'=B_\perp/\gamma_w$. However, closer to the light cylinder the radial component can still be a sizeable fraction of the total field as $B_r/B_\phi\sim R^{-1}$. This is especially the case at higher latitudes, or where the magnetosphere has been opened and the magnetic field distorted by flaring activity, in which case the radial magnetic field can be significantly increased \citep{2020ApJ...896..142B,2023MNRAS.524.6024S}.

 The required magnetic field is found at radii of $R_{FRB}=R_0\sim2.46\times 10^{11}bl\nu_9^{-1}\mu_{33}P^{-2}\text{ cm}$, providing a direct value for the emission radius. Here, $R_0$ denotes the emission radius assuming an unperturbed magnetosphere and wind. In the case of a change to the magnetic configuration of the magnetar, $R_{FRB}$ will increase to values larger than $R_0$ as $B$ will decrease more slowly with radius due to the opened field lines.

\subsection{Time-scale}
The second constraint can be derived by examining the time-scale $\delta t$ of observed FRBs, which is typically on the order of a few milliseconds \citep[e.g.][]{Zhang2020}. For a spherical or conical shell moving relativistically towards the observer, the intrinsic time-scale is 

\begin{equation}
    \delta t \approx \frac{R_{FRB}}{2 c \gamma_w^2}.
\end{equation}
This condition imposes an upper limit on the FRB radius: $R_{FRB} \lesssim 6 \times 10^7 \gamma_w^2\delta t_{-3}\text{ cm}$. Therefore, in order for FRB emission to occur in the magnetar wind ($R_{FRB} > R_{LC}= 4.8\times10^9 P \text{ cm}$), the emitting region must be moving relativistically towards the observer. 

\subsection{Shell size}
The final constraint is derived from the condition that the SME must occur over a distance $d$ that is small compared to the coherent length of the magnetic field. This is required so that the background field and hence the emission frequency do not substantially change in the masing region. As $B\propto R^{-1}$ in the magnetar wind, we impose the constraint $d<R$ as an upper limit on the width of the shell where SME occurs.

To estimate the required size of the emitting region, consider that the isotropic equivalent FRB energy, $E_{FRB}$, is emitted from a spherical shell with width $d$ and inner radius $R_{FRB}$. The volume of this region is $V' =  N/n' = \frac{4 \pi}{3}((R_{FRB}+d)^3-R_{FRB}^3)$, with the total number of particles given by $N = 6.1\times 10^{44} E_{FRB,39} \gamma_{w}^{-1}(\gamma_{av}-1)^{-1}f_{inv}^{-1}$. Here $f_{inv}$ is the fraction of particle energy that contributes to the SME \citep{2025MNRAS.538.1029L} and $\gamma_{av}$ is the average Lorentz factor of the particles. The number density at $R_{FRB}$ in the wind frame is $n'\sim3.1\times10^{9}w^{-1}\nu_9^2\gamma_w^{-2}l^{-2}\sigma^{-1}\text{ cm}^{-3}$. The volume of the emitting region is therefore 
\begin{equation}
    V' = 1.97\times10^{35} \frac{E_{FRB,39}\gamma_{w}\sigma l^2w }{\nu_9^2(\gamma_{av}-1)f_{inv}} \text{ cm}^{3}.
\end{equation}
Using the expression for $R_{FRB}$, the fractional shell width $d/R_{FRB}$ is given by

\begin{equation}
 \frac{d}{R_{FRB}}  =\left(3.15\frac{E_{FRB,39}\gamma_w\sigma\nu_9P^6 w}{lb^3\mu_{33}^3(\gamma_{av}-1)f_{inv}}+1\right)^{1/3}-1.
 \label{eq:dwind}
\end{equation}
It can be seen that the constraint $d/R_{FRB}<1$ is satisfied most easily for low periods and high magnetic fields, as both of these result in larger values of $R_{FRB}$. In particular, the shell width is so sensitive to the period due to its impact on the light cylinder radius.



\begin{figure*}
	\centering 
       \captionsetup{width=2\columnwidth}
{\phantomsubcaption\label{fig:lp}%
\phantomsubcaption\label{fig:fs}%
    }
	\includegraphics[width = 2\columnwidth]{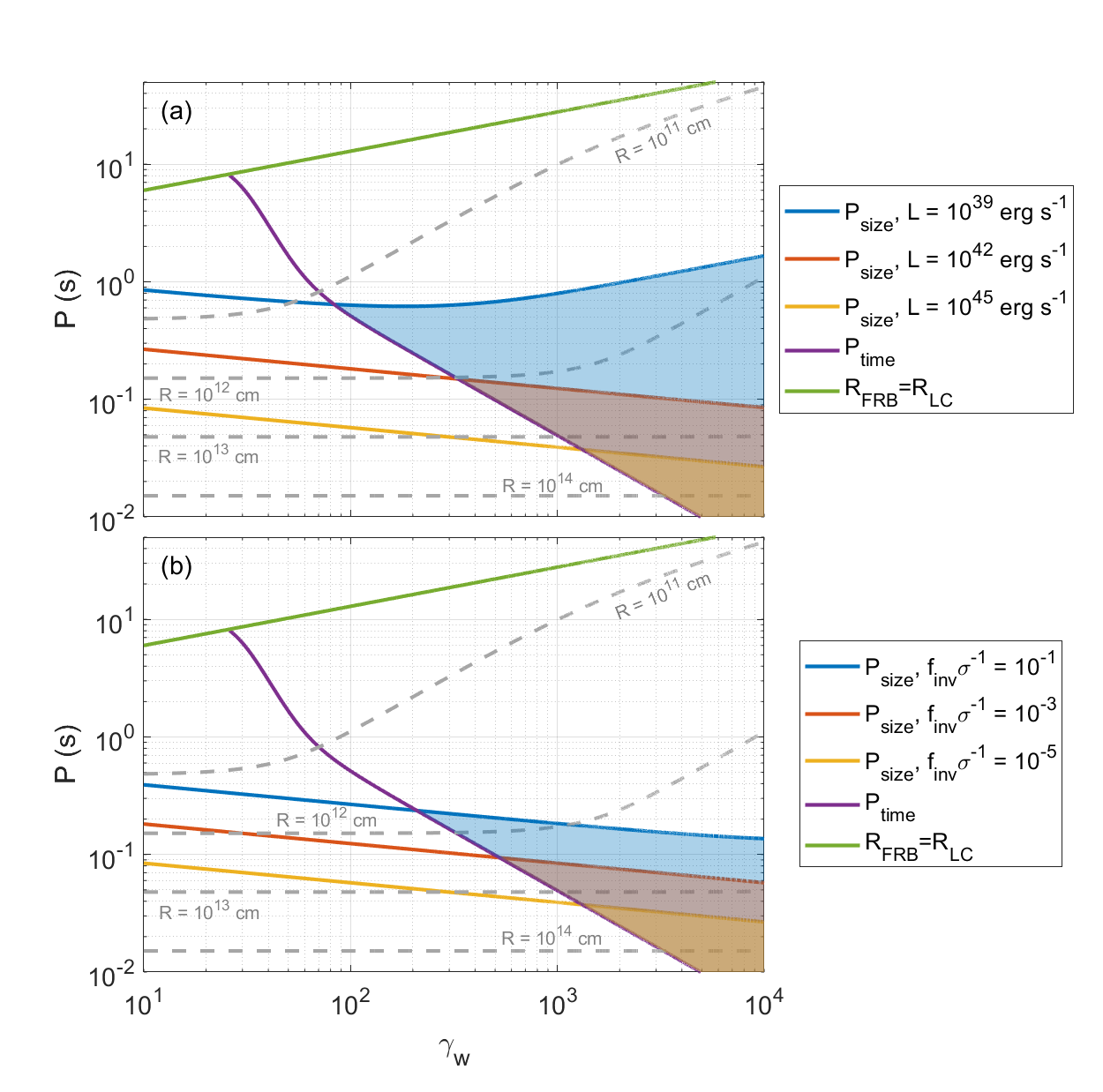}
	\caption{The $P-\gamma_w$ parameter space for the range of FRB luminosities. Panel (a) shows the parameter space for varying luminosities, with the solid blue, orange and yellow lines showing the upper limits from $P_{size}$ for $L_{FRB} = 10^{39} \text{ erg s}^{-1}$, $L_{FRB} = 10^{42} \text{ erg s}^{-1}$ and $L_{FRB} = 10^{45} \text{ erg s}^{-1}$ respectively. Panel (b) displays parameter space for varying values of $f_{inv}\sigma^{-1}$. In this case, the solid blue, orange and yellow lines show the upper limits from $P_{size}$ for $f_{inv}\sigma^{-1} = 10^{-1}$, $f_{inv}\sigma^{-1} = 10^{-3}$ and $f_{inv}\sigma^{-1} = 10^{-5}$ respectively. As $P_{time}$ is independent of $L_{FRB}$ in both panels, it is shown as a purple line which is valid for all parameters. The shaded region corresponds to the allowed parameter space where an FRB can be produced, with the colours corresponding to the different luminosity values in panel (a), and the different values of $f_{inv}\sigma^{-1}$ in panel (b). Constant emission radii $R$ are shown as dashed grey lines, and the upper limit from the condition $R_{FRB}>R_{LC}$ is shown in green. In all cases $\theta = 10^{-2}$, $\nu = 1\text{ GHz}$, $\delta t = 1 \text{ ms}$, $\mu = 10^{33}\text{ G cm}^{3}$ and $l = 0.6$. In panel (a), $f_{inv}\sigma^{-1} = 10^{-2}$, while in panel (b) the luminosity is $L_{FRB} = 10^{42} \text{ erg s}^{-1}$. }
	\label{fig:par_plots}
\end{figure*}

\subsection{Examining the parameter space}
\label{sec:par}
To investigate the allowed physical conditions for this model using the above constraints, we examine the $P-\gamma_w$ parameter space. Expressing the constraints as limits on the magnetar period results in

\begin{align}
    P_{size}&\lesssim 0.53(b\mu_{33})^{1/2}\left(\frac{f_{inv,-2}l(\gamma_{av}-1)}{E_{FRB,39}\gamma_w\sigma\nu_9w}\right)^{1/6}\text{ s},\\
    P_{time} & \gtrsim\frac{63.6}{\gamma_w}\left(\frac{bl \mu_{33}}{\nu_9\delta t_{-3}}\right)^{1/2}\text{ s},
\end{align}
where $P_{size}$ is the maximum period allowed by the shell size constraint, and $P_{time}$ is the minimum period allowed by the time-scale constraint. As expected due to the time-scale requirements, FRB emission outside the light cylinder requires quite high Lorentz factors. The minimum Lorentz factor which satisfies both conditions is $\gamma_w \gtrsim 313 l^{2/5}E_{FRB,39}^{1/5}w^{1/5}\sigma^{1/5}f_{inv,-2}^{-1/5}(\gamma_{av}-1)^{-1/5}\delta t_{-3}^{-3/5}\nu_9^{-2/5}$, emphasising the need for a relativistic magnetar wind.

Quantities affecting the number of particles required only have an impact on the upper limit $P_{size}$. Increasing $E_{FRB}$ (or the equivalent luminosity $L_{FRB} \sim E_{FRB}/\delta t$) and $\sigma$, or decreasing $f_{inv}$ and $\gamma_{av}$ all result in a more restrictive limit due to the need for more particles to produce the FRB. However, the dependence on all of these quantities except $\gamma_{av}$ (and hence the temperature) is very weak. The value of $\gamma_{av}$ is calculated by integrating over the distribution function, $\gamma_{av}=\int d^3\mathbf{p}\gamma(\mathbf{p})F(\mathbf{p},\eta=1)/\int d^3\mathbf{p}F(\mathbf{p},\eta=1)$. Although the value of $\gamma_{av}$ for the initial thermal distribution is very low for the temperatures in question (e.g. $\gamma_{av}\sim1.025$ for $\theta=10^{-2}$), the non-resonant interactions dramatically increase the average kinetic energy, resulting in values of $\gamma_{av}\gtrsim1.3$ even for initial temperatures of $\theta=10^{-3}$. This increase is clear by comparing the initial and final distributions in Fig. \ref{fig:disa} and \ref{fig:disb} respectively. It should be noted that $l$ also has a temperature and magnetisation dependence, though the range of allowed values is much smaller than that of other terms, as shown in Section \ref{sec:grcalc}. As a result we keep $l=0.6$ (appropriate for O-mode emission) as an approximate value throughout.

\begin{figure}
	\centering 
       \captionsetup{width=\columnwidth}

	\includegraphics[width = \columnwidth]{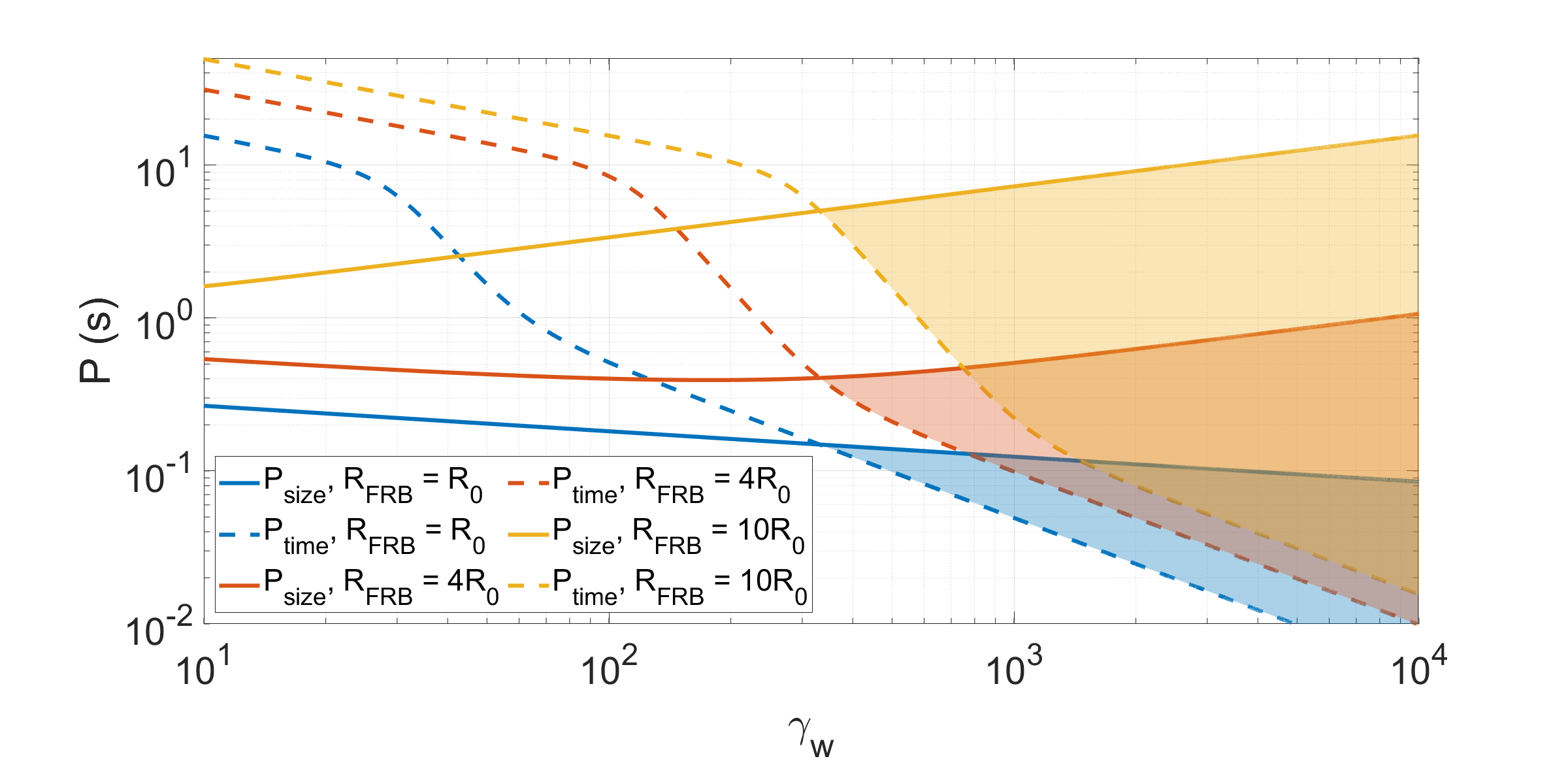}
	\caption{The $P-\gamma_w$ parameter space for different magnetic field configurations, showing how the allowed period depends strongly on the neutron star magnetic field. The solid blue, orange and yellow lines show the upper limits from $P_{size}$ for $R_{FRB} = R_0$, $R_{FRB} = 4R_0$ and $R_{FRB} = 10R_0$ respectively. The equivalent lower limits for $P_{time}$ are shown as dashed lines of the same colours. The shaded region corresponds to the allowed parameter space where an FRB can be produced, with the colours corresponding to the different values of $R_{FRB}$. In all cases $\theta = 10^{-2}$, $\nu = 1\text{ GHz}$, $\delta t = 1 \text{ ms}$, $\mu = 10^{33}\text{ G cm}^{3}$, $L_{FRB} = 10^{42} \text{ erg s}^{-1}$, $f_{inv}\sigma^{-1} = 10^{-2}$ and $l = 0.6$.}
	\label{fig:rv}
\end{figure}

Both limits are proportional to $(b\mu)^{1/2}$, implying that magnetars with higher magnetic fields also need correspondingly higher periods to allow a similarly sized region of parameter space for FRB emission. As the period increases, the light cylinder radius expands, resulting in a lower value of $B$ in the wind, counteracting the influence of the increased surface field. Varying these parameters does not change the minimum $\gamma_w$ required. The time-scale constraint is naturally more sensitive to both frequency and burst duration, with the limit becoming more constraining as both parameters decrease. As the time-scale is independent of the energetics of the burst, it does not depend on the luminosity or efficiency. The impact of temperature and magnetisation is only through the effect of these factors on the value of $\omega_m$ and hence on $R_{FRB}$. In general, $P_{time}$ is much more sensitive to variations in the parameters due to the difference in exponent between the two limits. 

The allowed parameter space is shown for various scenarios in Figs. \ref{fig:par_plots} and \ref{fig:rv}. In all cases $b$ is calculated assuming that $\xi=R_{0}/R_{LC}$, i.e. that $B_\phi=B_r$ at the light cylinder. This is an approximation, as the true value will depend on both the latitude and the exact configuration of the magnetic field. In Fig. \ref{fig:lp} the parameter space is displayed for a range of luminosities based on FRB observations \citep{ 2019Natur.572..352R,2020Natur.587...59B,  2023Sci...382..294R} for $\theta = 10^{-2}$, $f_{inv} = 10^{-2}$ (taken as a typical value based on results in \cite{2025MNRAS.538.1029L}), $\sigma =1$, $l = 0.6$, $\delta t = 1 \text{ ms}$ and $\nu = 1\text{ GHz}$. Fig. \ref{fig:fs} uses the same parameters with the FRB luminosity fixed at $L_{FRB} = 10^{42}\text{ erg s}^{-1}$. In both these figures $R_{FRB} = R_0$, i.e. an unperturbed magnetosphere is assumed. Also shown is the upper limit imposed by the condition $R_{FRB}>R_{LC}$, though this is less constraining than $P_{size}$ for all the parameters displayed. The regions of the parameter space where $b\neq1$ can be seen where the lines of constant radii (dashed grey lines) deviate from the horizontal. As expected, this occurs at high Lorentz factors and smaller radii.

From Fig. \ref{fig:lp} it is clear that weaker bursts have a larger allowed parameter space, extending to both lower Lorentz factors and longer periods. The lower limit is not affected by the change in luminosity as the time-scale is independent of $L_{FRB}$. Emission radii of $R_{FRB}\sim10^{13-14}\text{ cm}$ are favoured for typical FRB luminosities of $L_{FRB}=10^{42}\text{ erg s}^{-1}$, with the radius increasing as the luminosity increases due to the need for more particles to provide the free energy. Fig. \ref{fig:fs} shows the impact of changing the efficiency of the SME or the magnetisation of the wind. The quantities have been combined for the purposes of presentation, as increasing $f_{inv}$ has the same effect as decreasing $\sigma$, with both changing the required volume of particles in the emission region. As expected, decreased efficiency or equivalently increased magnetisation reduces the allowed parameter space by imposing a more constraining upper limit from $P_{size}$. As in the case of varying $L_{FRB}$, only the upper limit is affected by the change in $f_{inv}\sigma^{-1}$.

Both cases demonstrate that periods of $P\lesssim0.1\text{ s}$ are favoured except for all but the weakest of FRBs, provided that the magnetosphere is unperturbed. In contrast, Fig. \ref{fig:rv} shows the impact of a distortion to the magnetosphere resulting in increased magnetic field outside the light cylinder, as may be expected both before and after large flares \citep{2020ApJ...896..142B,2023MNRAS.524.6024S}. To demonstrate this, the allowed parameter space is shown for $R_{FRB} = R_0$, $R_{FRB} = 4R_0$ and $R_{FRB} = 10R_0$. All other parameters are the same as in Fig. \ref{fig:fs} with $f_{inv}\sigma^{-1} = 10^{-2}$. As both limits contain a factor of $b^{1/2}$ the minimum Lorentz factor is not affected by the change in radius. However, the range of allowed periods shifts to higher values for larger radii, with even moderate increases bringing the preferred period range to the order of seconds.

\subsection{Emission escape}

A further crucial factor when considering FRBs is whether the signal can escape the emission environment without significant damping. Due to the extreme brightness of the FRB signal, the wave strength parameter $a_0=e E_0/(m_e \omega c )$ may be large ($a_0>1$), depending on the distance of the emission region from the magnetar \citep{2014ApJ...785L..26L}. Here, $E_0$ is the electric field of the wave. In this regime, non-linear effects may result in compression of the wind plasma, taking significant amounts of energy from the radio waves \citep[e.g.][]{2021ApJ...922L...7B, 2023ApJ...957..102G, 2024A&A...690A.332S}. However, the details of this mechanism are still uncertain, so we do not impose a hard limit on the allowed parameter space. We note that different authors derive different lower limits on the value of $R$ beyond which FRB emission can escape, in the range $R_{damp}\gtrsim 10^{11-12}\text{ cm}$, depending on the parameters of the FRB \citep{2024A&A...690A.332S, 2025arXiv250316054B}. In general, these values will not tighten the constraints presented in Section \ref{sec:par} except to lower the upper limits on some of the weaker FRB cases with the largest allowed parameter spaces. 

\section{Application to FRB 20200428}
\label{sec:frb}
The limits obtained in Section \ref{Sec:constraints} can be applied to FRB 20200428 since the properties of its associated progenitor, the galactic magnetar SGR 1935+2154, are known \citep{2020Natur.587...54C, 2020Natur.587...59B,2020ApJ...898L..29M,2021NatAs...5..378L}. This FRB is extremely weak by typical standards, with a luminosity of $L_{FRB}\sim3.6\times 10^{38}\text{erg s}^{-1}$ \citep{2020Natur.587...54C, 2020Natur.587...59B}. Here, we have taken the inferred luminosity of the STARE2 detection as it was the more luminous burst \citep{2020Natur.587...59B}.

\begin{figure}
	\centering 
    \captionsetup{width=\columnwidth}

	\includegraphics[width = \columnwidth]{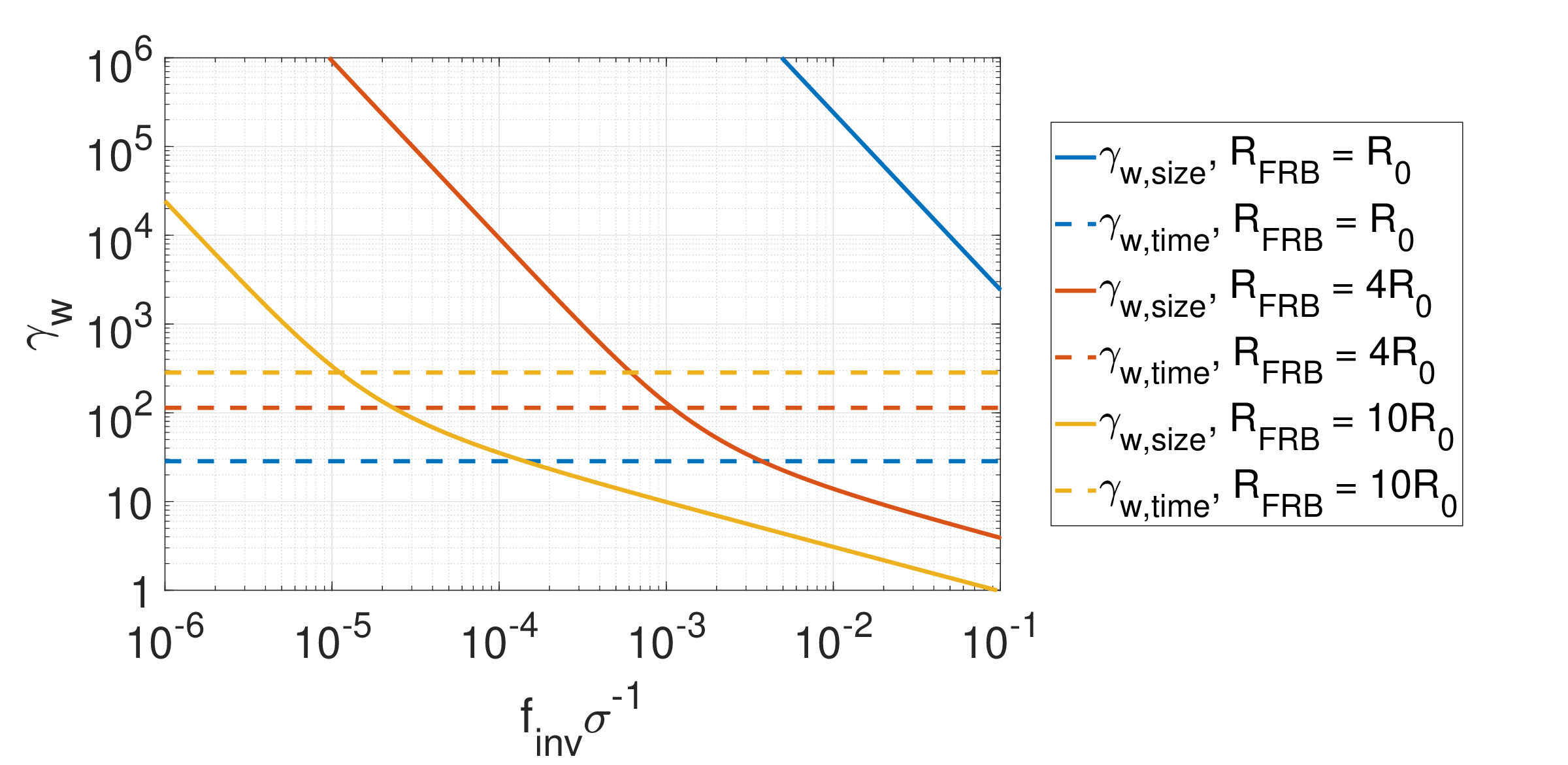}
	\caption{The $\theta-\gamma_w$ parameter space for FRB 20200428. The solid blue, orange and yellow lines show the lower limits  $\gamma_{w,size}$ for $R_{FRB} = R_0$, $R_{FRB} = 4R_0$ and $R_{FRB} = 10R_0$ respectively. The equivalent lower limits for $\gamma_{w,time}$ are shown as dashed lines of the same colours. In all cases $\theta=10^{-2}$ and $l=0.6$.}
	\label{fig:frb20}
\end{figure}

Using the magnetar properties of $P= 3.25\text{ s}$ and $B_* = 2.2\times10^{14}\text{ G}$ \citep{2014ApJS..212....6O, 2016MNRAS.457.3448I}, we calculate the lower limits imposed on $\gamma_w$ by the shell size and time-scale constraints.
These lower limits are plotted in Fig. \ref{fig:frb20} as a function of $f_{inv}\sigma^{-1}$ for $R_{FRB} = R_0$, $R_{FRB} = 4R_0$ and $R_{FRB} = 10R_0$ with $\theta=10^{-2}$ and $l=0.6$. For the unperturbed magnetosphere case (shown by the blue lines in Fig. \ref{fig:frb20}), the shell size constraint is more restrictive for all magnetisations and efficiencies. For the majority of the parameter space the lower limit is extremely high, with $\gamma_w>10^4$ even for $f_{inv}\sigma^{-1}=0.05$. This likely rules out a completely unperturbed scenario. As $R_{FRB}/R_0$ increases the allowed parameter space extends to lower values of $\gamma_w$ and $f_{inv}\sigma^{-1}$, with the crossover point where $\gamma_{w,size}=\gamma_{w,time}$ decreasing such that the time-scale constraint becomes the dominant limit for $R_{FRB}=10R_0$ for even low values of $f_{inv}\sigma^{-1}\gtrsim10^{-5}$. These results suggest two possibilities for FRB 20200428 in the context of the model presented in this work. In the first case, if the magnetosphere is unperturbed, the emission must originate from an extremely relativistic wind with $\gamma_w>10^4$ and relatively low magnetisation. On the other hand, if the magnetic field configuration has been altered, emission is possible for a wind across a much more achievable range of efficiencies, magnetisations and $\gamma_w$. 

This model does not intrinsically produce the observed X-ray emission associated with FRB 20200428 \citep[e.g][]{2020ApJ...898L..29M,2021NatAs...5..378L}. However, the majority of X-ray bursts from SGR 1935+2154 have been observed to not have associated radio emission \citep[e.g][]{2020Natur.587...63L}, suggesting the emission mechanism may not be shared. Furthermore, the slight timing differences between the X-ray and radio imply that the emission may occur in different locations, even if the initial source of the energy is the same perturbation of the magnetar \citep{2023ApJ...953...67G}. Previous theoretical works have also demonstrated that the X-ray burst may be emitted within the magnetosphere while the FRB itself is produced in the wind \citep[e.g.][]{2020ApJ...900L..26W,2021MNRAS.500.2704Y}. For these reasons we do not consider the lack of X-ray emission from the non-resonant Alfv\'en wave interaction to be a serious issue in this case. 
\section{Discussion}
\label{sec:dis}
\subsection{FRB and magnetar populations}
The allowed parameter space determined in Section \ref{Sec:constraints} can be compared with the known properties of the magnetar population. Observed magnetars have periods ranging from $P\sim2-12\text{ s}$, with a mean value of $P= 6.3\text{ s}$. Their surface magnetic fields are $B_*\sim10^{14}-2\times 10^{15}\text{ G}$, with a mean value of $B_* =3.4\times 10^{14}\text { G}$ \citep{2014ApJS..212....6O}.  Magnetars may also be born with much shorter periods on the order of milliseconds and even larger magnetic fields \citep[e.g. ][]{2019MNRAS.487.1426B, 2022ASSL..465..245D}. As can be seen in Figs. \ref{fig:lp} and \ref{fig:fs}, the optimal periods found for an unperturbed magnetosphere are at least an order of magnitude lower than the observed magnetar periods. This suggests that either this mechanism may produce FRBs in the environment of younger magnetars with shorter periods, or that flaring activity results in larger emission radii, as shown in Fig. \ref{fig:rv}. These younger magnetars may be necessary to support the energy budget required for very active repeaters \citep{2024ARNPS..74...89Z}. However, as young magnetars spin down very rapidly due to their huge magnetic fields \citep{2019MNRAS.487.1426B}, they may struggle to reproduce repeating FRBs over longer time-scales.

\begin{figure} 
	\centering 
    \captionsetup{width=\columnwidth}

	\includegraphics[width = \columnwidth]{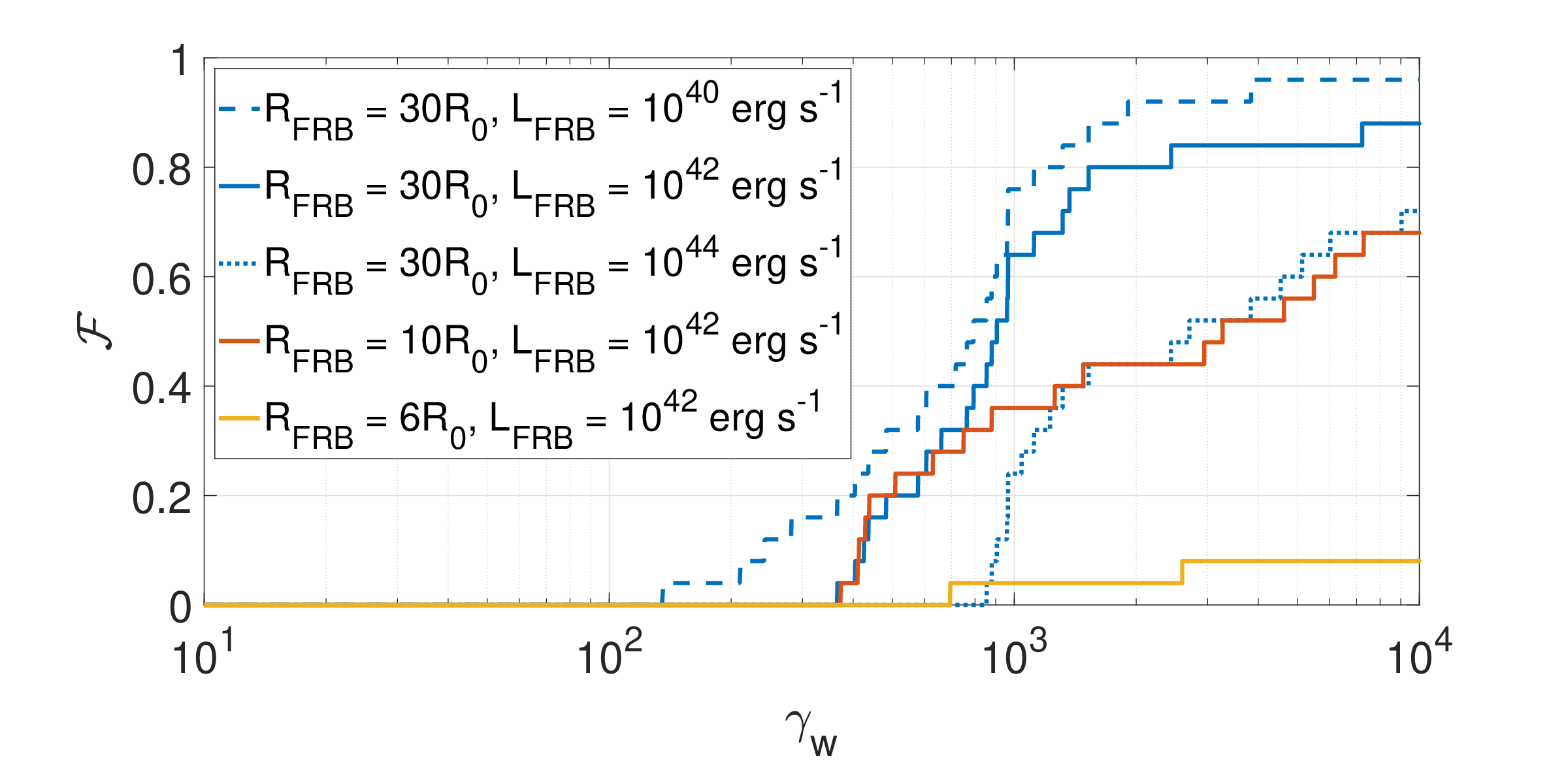}
	\caption{The fraction $\mathcal{F}$ of the observed magnetar population that can support FRB emission for a given wind Lorentz factor $\gamma_w$. The blue lines denote $R_{FRB}=30R_0$, with the dashed line denoting a luminosity of $L_{FRB}=10^{40}\text{ erg s}^{-1}$, the solid line showing $L_{FRB}=10^{42}\text{ erg s}^{-1}$ and the dotted line showing $L_{FRB}=10^{44}\text{ erg s}^{-1}$. The orange and yellow lines show $\mathcal{F}$ for $R_{FRB}=10R_0$ and $R_{FRB}=6R_0$ respectively, in both cases with $L_{FRB}=10^{42}\text{ erg s}^{-1}$. All other quantities have their fiducial values as in Fig. \ref{fig:lp}. }
	\label{fig:pop}
\end{figure}

The fraction $\mathcal{F}$ of the observed magnetar population that satisfies the conditions to support FRB emission for various $\gamma_w$ is shown in Fig. \ref{fig:pop} below for different $R_{FRB}/R_0$, assuming the same parameters as used in Fig. \ref{fig:rv}. Here, only magnetars with estimates of both $P$ and $B_*$ in the McGill Online Magnetar Catalog\footnote{\url{http://www.physics.mcgill.ca/~pulsar/magnetar/main.html}} \citep{2014ApJS..212....6O} have been used. The impact of different luminosities is demonstrated by the dashed, solid and dotted blue lines, which show $L_{FRB}=10^{40}\text{ erg s}^{-1}$, $L_{FRB}=10^{42}\text{ erg s}^{-1}$ and $L_{FRB}=10^{44}\text{ erg s}^{-1}$ respectively. As the allowed parameter space increases with $\gamma_w$ (see e.g. Fig. \ref{fig:lp}), so does $\mathcal{F}$, although significant fractions of $\mathcal{F}\gtrsim0.1$ are available close to the minimum $\gamma_w$ for both $R_{FRB}/R_0=6$ and $R_{FRB}/R_0=10$. It is apparent that significant changes to the initial unperturbed magnetic field are necessary for the current observed magnetar population to explain FRBs through this model. However, such changes are expected both in the form of enhanced pre-flare winds \citep{2020ApJ...896..142B} and from the impact of the flares themselves \citep[e.g.][]{2023MNRAS.524.6024S}. We note that these values of $\mathcal{F}$ are determined for optimal parameters (except for $L_{FRB}$ and $\theta$), and will decrease for less efficient maser emission or higher magnetisations.

 These results can be compared to the FRB rate, which is $\mathcal{R}_{FRB}\sim8\times10^{4}\text{ Gpc}^{-3}\text{yr}^{-1}$  for $E_{FRB}>10^{39}\text{ erg}$, with the rate at lower energies poorly constrained \citep{2022MNRAS.509.4775J, 2023ApJ...944..105S}. This is $\sim1\%$ of the magnetar rate, where we have taken a core-collapse supernova rate of $\mathcal{O}(10^4)\text{ Gpc}^{-3}\text{yr}^{-1}$ \citep{2019RPPh...82l6901O,2023ApJ...944..105S} and a typical magnetar lifetime estimate of $\sim 10^{4}\text{ yr}$ \citep{2017ARA&A..55..261K, 2019RPPh...82l6901O}. Furthermore, we have assumed that magnetars make up $\sim10\%$ of the neutron star population \citep[e.g. ][]{2007MNRAS.381...52G,  2019RPPh...82l6901O,2025ApJ...986...88S}. Therefore, in this scenario the values of $\mathcal{F}$ obtained in Fig. \ref{fig:pop} are sufficiently high to explain the observed FRB population. We note that both the typical magnetar lifetime and the fraction of neutron stars that are magnetars are uncertain.

The difficulty of obtaining extremely high luminosity bursts ($L_{FRB}\gtrsim10^{44}\text{ erg s}^{-1}$) which require very high $\gamma_w$ is tempered by their relative rarity. Models of the FRB energy distribution show an upper cutoff on the energy of $E_{FRB}\sim10^{41-42}\text{ erg}$, with a relative rate of $\mathcal{R}_{FRB}(E_{FRB}=10^{42}\text{ erg})/\mathcal{R}_{FRB}(E_{FRB}=10^{39}\text{ erg}) \lesssim10^{-4}$, suggesting only a very small fraction of FRBs have such a high energy and luminosity \citep{2022MNRAS.511.1961H, 2022MNRAS.509.4775J, 2023ApJ...944..105S}.

As the energy required to produce an FRB is reduced from its isotropic equivalent value due to relativistic effects, the energy budget required to produce FRBs is readily satisfied by typical starquakes. For the values of $\gamma_w$ found in Section 3, the true energy required to produce an FRB with isotropic energy of $E_{FRB}=10^{39} \text{erg}$ is reduced by a factor of $\sim10^{5-6}$ in this model. Therefore, a typical starquake with $E_Q\sim 10^{42}\text{ erg}$ \citep[e.g.][]{2025arXiv250812567Q} provides sufficient energy even when taking into account the losses due to the transmission of energy into the Alfv\'en waves and the efficiency of the maser mechanism. In the case of rapidly repeating bursts, which can have rates of up to $\sim 700 \text{ hr}^{-1}$ \citep[e.g][]{2025arXiv250714707Z}, the energy budget becomes more constraining. Even though these bursts typically have relatively low values of $E_{FRB}\sim 10^{37-39}\text{ erg}$ \citep{2022RAA....22l4002Z,2025ApJ...992..185W, 2025arXiv250714707Z}, such hyperactive FRBs would likely require either larger starquakes, which many have energies of up to $10^{44}\text {erg}$ \citep[e.g.][]{2020ApJ...900L..21Y}, or a chain of quakes in quick succession. However, we note that FRBs which such rapid rates of repetition are rare, with only a handful detected to date \citep[e.g.][]{2024MNRAS.534.3331K}.

\subsection{Properties of the SME}
The emission is produced by both the ordinary and extraordinary modes, depending on the plasma parameters, with the O-mode favoured at higher magnetisations and temperatures. In contrast, the slow extraordinary mode is favoured at lower magnetisations, with the growth rates of the two modes being roughly comparable at low temperatures. As the slow extraordinary mode is a trapped mode which needs to undergo mode conversion to escape, it is a less likely channel for FRB emission. However, observations of signals that are not strongly linearly polarized may indicate that the initial emission occurred in the low magnetisation regime, and that the resulting signal escaped after undergoing mode conversion. Although the general conditions for mode conversion are present, namely density and magnetic field gradients, both are over large scales, and a detailed investigation of the mechanism is outside the scope of this work. However, as there is a large part of the parameter space where the ordinary mode is dominant, this does not have a major impact on the overall viability of SME producing FRBs.

Both the O and X-modes are linearly polarized, supporting observations of high degrees of linear polarization in the majority of FRBs \citep{2023MNRAS.522.2448Q, 2023RvMP...95c5005Z}. However, the mechanism does not intrinsically produce the rapid swings of polarization angle seen in a small fraction of FRBs \citep{2024ApJ...972L..20N, 2025Natur.637...43M}. While this has led to arguments in favour of a magnetospheric origin of FRBs, it has been shown in the case of SME in both non-relativistic scenarios \citep{2012PhPl...19h2902W} and relativistic shocks \citep{2024PhRvL.132c5201I} that a change in emission mode and polarization angle is possible outside the magnetosphere due to the impact of waves (either the pervading Alfv\'en waves or the SME itself) on the emission process. The impact of the pervading Alfv\'en waves on the maser instability has also been shown to boost the O-mode growth rate in similar (non-relativistic) scenarios \citep{2012PhPl...19h2902W, 2013ApJ...770...75Z}. As we have not included the impact of these waves on the maser instability, this may result in a change in peak frequency for those parameters which currently have higher extraordinary mode growth rates. 

The results for the growth rate calculations shown in Section \ref{sec:grcalc} display considerable similarities to the results for the efficiency of SME from relativistic shocks, despite the different formation mechanisms for the population inversion. In the shock model, particles which gyrate around the enhanced magnetic field near the shock front form a ring distribution and produce SME \cite[e.g.][]{1988PhFl...31..839A, 1992ApJ...391...73G, 2006ApJ...653..325A, Plotnikov2019}. In particle-in-cell (PIC) simulations efficiencies of $\sim10^{-3}\sigma^{-1}$ for $\sigma>>1$ were found in such scenarios \citep{Sironi2021}, showing a similar dependence on $\sigma$ as for the crescent-shaped distribution discussed in this work. Furthermore, the efficiency also drops significantly at temperatures of $\theta\gtrsim10^{-1.5}$ \citep{2020MNRAS.499.2884B}, similar to the maximum temperature of $\theta\sim0.02$ found above. We note that the values of $\Gamma_i$ and the efficiency values reported in the PIC simulations are not the same physical quantity, as the efficiency is the fraction of total energy emitted by the maser rather than the growth rate. However, we expect the same trends to hold for both.

In order for SME to efficiently extract energy from the population inversion, it must occur on a shorter time-scale than other plasma instabilities. To obtain estimates for the stability of the population distribution, we examine the case of a bi-Maxwellian distribution with the same plasma parameters as the crescent shaped distribution \citep{2025MNRAS.538.1029L}. Such distributions are susceptible to the firehose and mirror instabilities \citep{gary1993theory}. While the crescent and bi-Maxwellian distributions are not directly equivalent, the comparison provides an approximation for the stability criteria. For bi-Maxwellian distributions, these are $A>1-1/\beta_\parallel$ for the firehose instability and $A<1+1/\beta_\parallel$ for the mirror instability \citep{gary1993theory}. Here, $A=\theta_\perp/\theta_\parallel$ is the temperature anisotropy and $\beta_\parallel=2\theta_\parallel/\sigma$ is the parallel plasma beta. As this model is only viable in the highly magnetised and lower temperature region of the parameter space (see Sec. \ref{sec:maser}), extreme anisotropies are required to violate these criteria. Therefore, the distribution should be stable against both instabilities in the regime of interest. However, this would not be the case for scenarios with lower magnetisations or higher temperatures.

Energy must also be extracted sufficiently quickly in comparison to the time-scale of the system. The typical duration of the Alfv\'en wave packet released by the original starquake is $t\sim10^{-3}-10^{-2}\text{ s}$ \citep[e.g][]{2020ApJ...900L..21Y}, depending on the properties of both the magnetar and the quake itself. In contrast, the typical maser time-scale is $t_M\sim \Gamma_i^{-1} \sim (10^{-4}-10^{-5})\Omega_8 \text{ s}$. Therefore, there is sufficient time available for SME to extract a significant amount of energy from the inverse population, though the time-scale can become more constraining at higher magnetisations.

\subsection{Other observational constraints}

A further observational constraint for at least one FRB (FRB 20221022A) comes in the form of scintillation measurements \citep{2025Natur.637...48N}. The upper limit on the size of the emitting region in this case is $R_{FRB}\sim (0.8-5)\times10^{11}\text{ cm}$, where we have used the upper bound on the lateral size of the emission region as obtained by \citet{2025Natur.637...48N}. For this particular FRB, this value is quite constraining, and requires very high efficiencies and low magnetisations to satisfy. However, we note that scattering may occur as a result of the signal's propagation through the relativistic magnetar wind, complicating the analysis \citep[e.g.][]{2022MNRAS.511.4766S}.

While both polarization swings and scintillation measurements impose tighter constraints on the magnetar wind scenario in comparison with magnetospheric models, importantly wind models avoid major difficulties raised by the potential damping of the FRB \citep[e.g.][]{2021ApJ...922L...7B, 2023ApJ...957..102G, 2024ApJ...975..223B}. We note that the importance and details of the damping mechanisms are still under debate \citep[e.g.][]{2022MNRAS.515.2020Q,2024MNRAS.529.2180L}. 

As the emission occurs in the magnetar wind, the size of the emission region may be large relative to the beaming angle of $\sim \gamma_w^{-1}$. In this case, the bandwidth of the signal is broadened by geometric effects, with $\Delta \nu /\nu \sim 0.58$ \citep{2024ApJ...974..160K}. This favours one-off FRBs, which typically have broader bandwidths, while repeated bursts tend to be narrower \citep[e.g.][]{2021ApJ...923....1P}. However, smaller scale structures in the wind itself may limit the size of the region where masing occurs, allowing narrower bandwidth bursts. This topic requires a more detailed analysis of both the wind structure and the nature of the Alfv\'en waves released by the initial starquake. \footnote{Alfv\'en waves may also undergo mode conversion to electromagnetic waves when $\omega>\omega_p$. However, this would require large magnetisations of $\sigma \sim 10^{10}\Omega_9^2\omega_{p,4}^{-2}$, assuming kHz Alfv\'en waves and SME at typical FRB frequencies. This process is therefore not applicable to this model.}
\section{Conclusions}
\label{sec:conclusions}
In this work, we have examined the allowed parameter space for FRBs emitted by SME in relativistic magnetar winds. We consider that the necessary population inversion has been formed by non-resonant interactions between Alfv\'en waves produced by the central magnetar and the wind. For all parameters a relativistic wind is required, with minimum Lorentz factors of $\gamma_w \gtrsim 310l^{2/5}E_{FRB,39}^{1/5}w^{1/5}\sigma^{1/5}f_{inv,-2}^{-1/5}(\gamma_{av}-1)^{-1/5}\delta t_{-3}^{-3/5}\nu_9^{-2/5}$. Provided that the magnetic field is perturbed by the flaring activity, the allowed periods align with the observed magnetar period distribution. In the case of an unperturbed magnetic field configuration, the typical period required is $P\lesssim0.1\text{ s}$, suggesting that only young magnetars may be valid sources of FRBs in this particular scenario. The emission takes place at typical distances of $R_{FRB}\sim 10^{13-15}\text{ cm}$, outside the light cylinder and at a sufficient distance to avoid expected damping mechanisms. This model therefore presents a robust explanation for the production of FRBs in magnetar winds and can be used to probe the physical conditions in the emission region, such as the wind Lorentz factor and magnetisation.

\section*{Acknowledgements}

The authors would like to thank the reviewer for their helpful and insightful comments.
AP acknowledges support from the European Union (EU) via ERC consolidator grant 773062 (O.M.J.). KL acknowledges the support of the Irish Research Council through grant number GOIPG/2017/1146 as well as funding obtained by the above ERC grant. 

\section*{Data Availability}
The code used to perform the calculations presented in this paper is available upon request.

\bibliographystyle{mnras}
\begingroup
\setlength{\emergencystretch}{3em} 
\sloppy
\raggedright                     
\small
\bibliography{paper_bibl}
\endgroup
\appendix

\section{Appendix A: Calculation of Growth Rates}
The dispersion equation can be written as $\text{det}(\mathbf{\Lambda})=0$, where 

\begin{equation}
    \Lambda_{ij} =  N_r^2\left(\frac{k_ik_j}{k^2}-\delta_{ij}\right)+\varepsilon_{ij}.
    \label{eq:aplambda}
\end{equation}
Here, $N_r=kc/\omega$ is the refractive index and $\varepsilon$ is the dielectric tensor given by \citep[e.g. ][]{Wu1985,1992wapl.book.....S}

\begin{equation}
    \varepsilon_{ij} = \delta_{ij} + \sum_sQ^s_{ij}(\mathbf{k},\omega),
\end{equation}
where 
\begin{align}
    \mathbf{Q}^s &= 2\pi\frac{\omega_{ps}^2}{\omega^2}\int_{-\infty}^{\infty}\text{d}u_{\parallel}\int_{0}^{\infty}\text{d}u_{\perp}\left\lbrace\frac{u_\parallel}{\gamma}\left(u_\perp\frac{\partial}{\partial u_\parallel}-u_\parallel\frac{\partial}{\partial u_\perp}\right)\right\rbrace \times \nonumber\\
    &\times F_s(u_\parallel,u_\perp)\mathbf{\hat{e}_z\hat{e}_z}+\omega\left[\frac{\partial}{\partial u_\perp}+\frac{k_\parallel}{\gamma\omega}\left(u_\perp\frac{\partial}{\partial u_\parallel}-u_\parallel\frac{\partial}{\partial u_\perp}\right)\right]\times     \label{eq:lambda}
 \\
    &\times F_s(u_\parallel,u_\perp) \sum_{n=-\infty}^{\infty}\frac{\mathbf{T_n}}{\gamma \omega - n \Omega_s - k_\parallel u_\parallel}. \nonumber
\end{align}
Here, the superscript $s$ denotes the species (e.g. electrons/positrons) of particles, $n$ is the harmonic number and $\mathbf{u}$ is the relativistic momentum per unit mass. We have assumed that $\mathbf B = B\mathbf{\hat{e}_z}$. The tensor $\mathbf{T_n}$ is given by

\begin{align}
   &\mathbf{T_n}(b) = \nonumber \\  
&\begin{pmatrix}
\frac{n^2 \Omega_s^2}{k_\perp^2}J_n^2(b) & -i\frac{n \Omega_s}{k_\perp}u_\perp J_n(b)J'_n(b) & \frac{n \Omega_s}{k_\perp}u_\parallel J_n^2(b)\\
i\frac{n \Omega_s}{k_\perp}u_\perp J_n(b)J'_n(b) & u_\perp^2J_n'^2(b) & i u_\perp u_\parallel J_n(b)J'_n(b)\\
\frac{n \Omega_s}{k_\perp}u_\parallel J_n^2(b) & -i u_\perp u_\parallel J_n(b)J'_n(b) & u_\parallel^2J_n^2(b)
\end{pmatrix},
\label{eq:Tn}
\end{align}
where $b = k_\perp u_\perp/\Omega_s$ and $J_n(b)$ is a Bessel function of the first kind, with $J'_n(b) = \frac{\partial J_n(b)}{\partial b}$.

To simplify the calculation of the peak masing frequency $\omega_m$ we make the approximation that plasma consists of a thermal component which dominates the dispersion relation (denoted by the subscript $0$) and that the component containing the population inversion (denoted by the subscript $e$) only contributes to the damping or amplification of the modes propagating through the background plasma. For a cold electron - proton plasma, the dispersion relation is given by $\text{Re}\left(\Lambda_0\left(\mathbf{k},\omega\right)\right)=0$ \citep[e.g. ][]{1969AdPlP...3....1B}, where

\begin{equation}
    \Lambda_0 = \begin{vmatrix}
        \varepsilon - N_r^2\cos^2\phi & -ig & N_r^2\cos\phi\sin\phi\\
        ig & \varepsilon - N_r^2 & 0\\
        N_r^2\cos\phi\sin\phi & 0 &h-N_r^2\sin^2\phi
    \end{vmatrix},
\end{equation}
where $\phi$ is the propagation angle. The quantities $\varepsilon$, $g$ and $h$ are given by 
\begin{align}
    \varepsilon &=1-\sum_s\frac{\omega_{ps}^2}{\omega_r^2-\Omega_s^2}, \nonumber \\
    g &=\sum_s\frac{\omega_{ps}^2\Omega_s}{\omega_r(\omega_r^2-\Omega_s^2)}, \\
    h &= 1-\sum_s\frac{\omega_{ps}^2}{\omega_r^2}. \nonumber
\end{align}
Due to mass symmetry, the $\varepsilon_{xy}$ and $\varepsilon_{yx}$ terms vanish for an electron-positron plasma \citep{1992JPlPh..47..295S,1993PhRvE..47..604I}. The resulting cold plasma dispersion relation is therefore

\begin{equation}
    \Lambda_0 = \begin{vmatrix}
        \varepsilon - N_r^2\cos^2\phi & 0 & N_r^2\cos\phi\sin\phi\\
        0 & \varepsilon - N_r^2 & 0\\
        N_r^2\cos\phi\sin\phi & 0 &h-N_r^2\sin^2\phi 
    \end{vmatrix} = 0.
\end{equation}

In either scenario, using the values for the real part of the wave frequency $\omega_r$, the growth rate $\omega_i$ can be approximated as \citep{Wu1985}

 \begin{equation}
     \frac{\omega_i}{\omega_r} = -\frac{ \text{Im}(\Lambda(\mathbf{k},\omega)}{\omega_r\frac{\partial}{\partial \omega_r}(\text{Re}(\Lambda_0(\mathbf{k},\omega_r))},
 \end{equation}
where we have assumed $\omega_i<<\omega_r$. To the lowest order in $n_e/n_o$, $\text{Im}(\Lambda(\mathbf{k},\omega))$ can be written as

\begin{align}
    \text{Im}(\Lambda(\mathbf{k},\omega)) &= \psi_{xx}\text{Im}(Q^e_{xx}(\mathbf{k},\omega_r))+\psi_{yy}\text{Im}(Q^e_{yy}(\mathbf{k},\omega_r))  \nonumber\\&+\psi_{zz}\text{Im}(Q^e_{zz}
    (\mathbf{k},\omega_r)) 
    +\psi_{xz}\text{Im}(Q^e_{xz}(\mathbf{k},\omega_r))\nonumber \\
    &+\psi_{xy}\text{Re}(Q^e_{xy}(\mathbf{k},\omega_r))+\psi_{yz}\text{Re}(Q^e_{yz}(\mathbf{k},\omega_r)),
\end{align}
with

\begin{align}
    \psi_{xx} &= N_r^4\sin^2\phi-N_r^2(\varepsilon\sin^2\phi+h) + \varepsilon h, \nonumber\\
    \psi_{yy}& = -N_r^2(\varepsilon\sin^2\phi+h\cos^2\phi)+\varepsilon h, \nonumber\\
    \psi_{zz} &= N_r^4\cos^2\phi-N_r^2\varepsilon(1+\cos^2\phi)+\varepsilon^2-g^2, \nonumber \\
    \psi_{xy} &=2g(h-N_r^2\sin^2\phi), \\
    \psi_{xz} &=2N_r^2(N_r^2-\varepsilon)\sin\phi\cos\phi, \nonumber\\
    \psi_{yz} &=2N_r^2g\sin\phi\cos\phi. \nonumber
\end{align}
As $g=0$ for the electron-positron plasma, $\psi_{xy} = \psi_{yz}=0$. Combining these equations and finally integrating along the contour defined by the resonance condition 

\begin{equation}
    \gamma \omega-n\Omega_s-k_\parallel u_\parallel = 0,
    \label{eq:rescon}
\end{equation}
allows the growth rate for a given mode to be calculated. Solving Eq. \ref{eq:rescon} provides the resonant perpendicular momentum,

\begin{equation}
\frac{u_n}{c}=\sqrt{\left(N_r\cos^2{\phi}-1\right)\frac{u_\parallel^2}{c^2}+2\frac{n \Omega_s}{\omega_r}N_r\cos{\phi}\frac{u_\parallel}{c}+\left(\frac{n^2\Omega_s^2}{\omega_r^2}-1\right)},
    \label{eq:resel}
\end{equation}
with the sum over all modes producing the results presented above in Section \ref{sec:maser}.

\subsection{Resonance Ellipses}
\begin{figure}
	\centering 
       \captionsetup{width=\columnwidth}
	\includegraphics[width = \columnwidth]{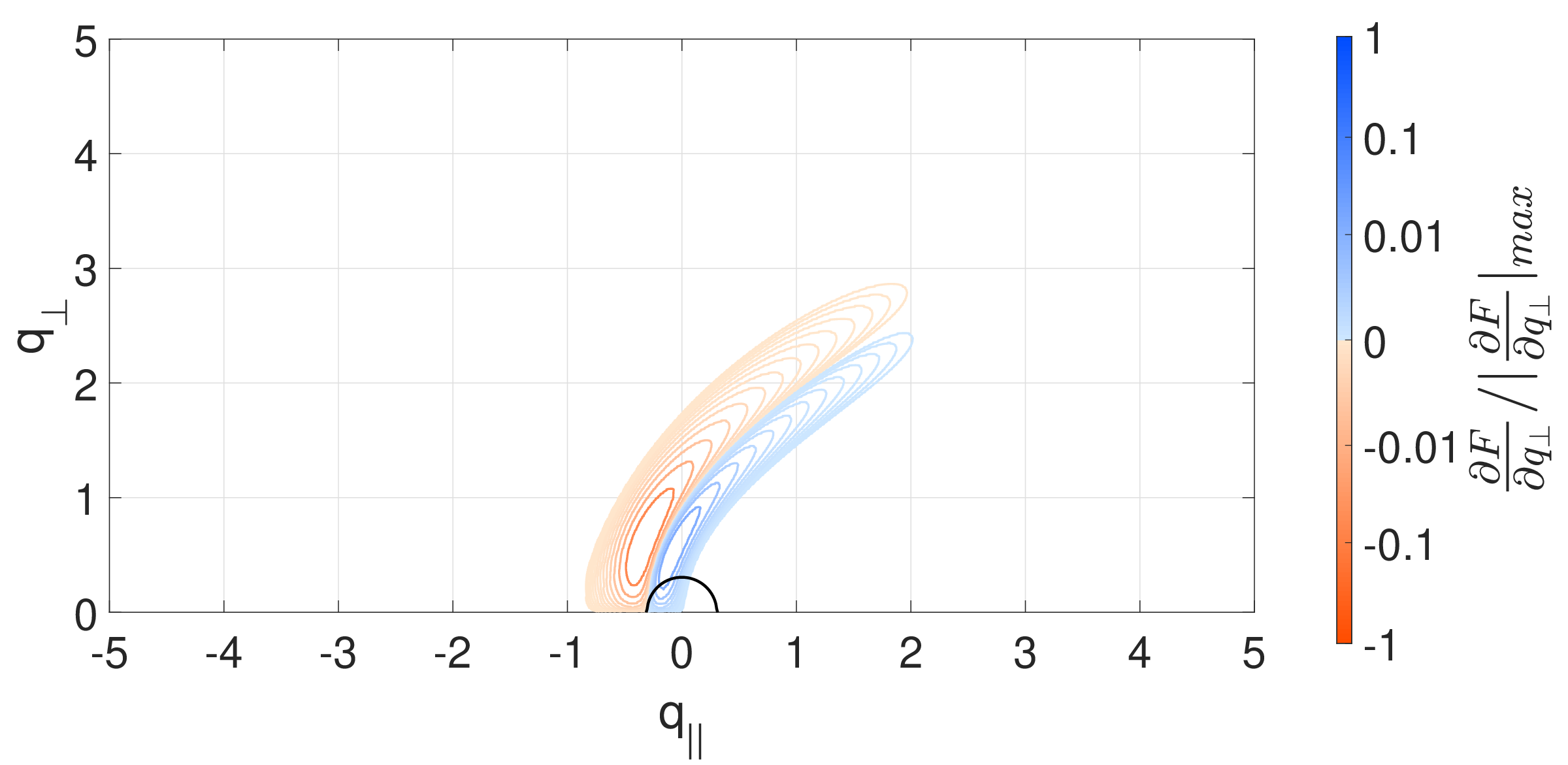}
	\caption{Contour plot showing $\frac{\partial F}{\partial q_\perp}$ for $\eta = 1$ and $\theta = 1/300$. The plot is normalised to $|\frac{\partial F}{\partial q_\perp}|_{max}$ for presentation purposes. Positive regions of the derivative are shown in blue, with negative regions in red. The black line shows the resonance ellipse for the slow X-mode with $\omega_r=0.95\Omega$ and $\phi = \pi/2$. This ellipse produces a positive growth rate at lower temperatures as it crosses more areas of  $\frac{\partial F}{\partial q_\perp}>0$. Note that this is not the only wavenumber to produce a positive growth, and has been chosen for visualisation purposes. }
	\label{fig:res_plota}
\end{figure}

\begin{figure}
	\centering 
       \captionsetup{width=\columnwidth}
	\includegraphics[width = \columnwidth]{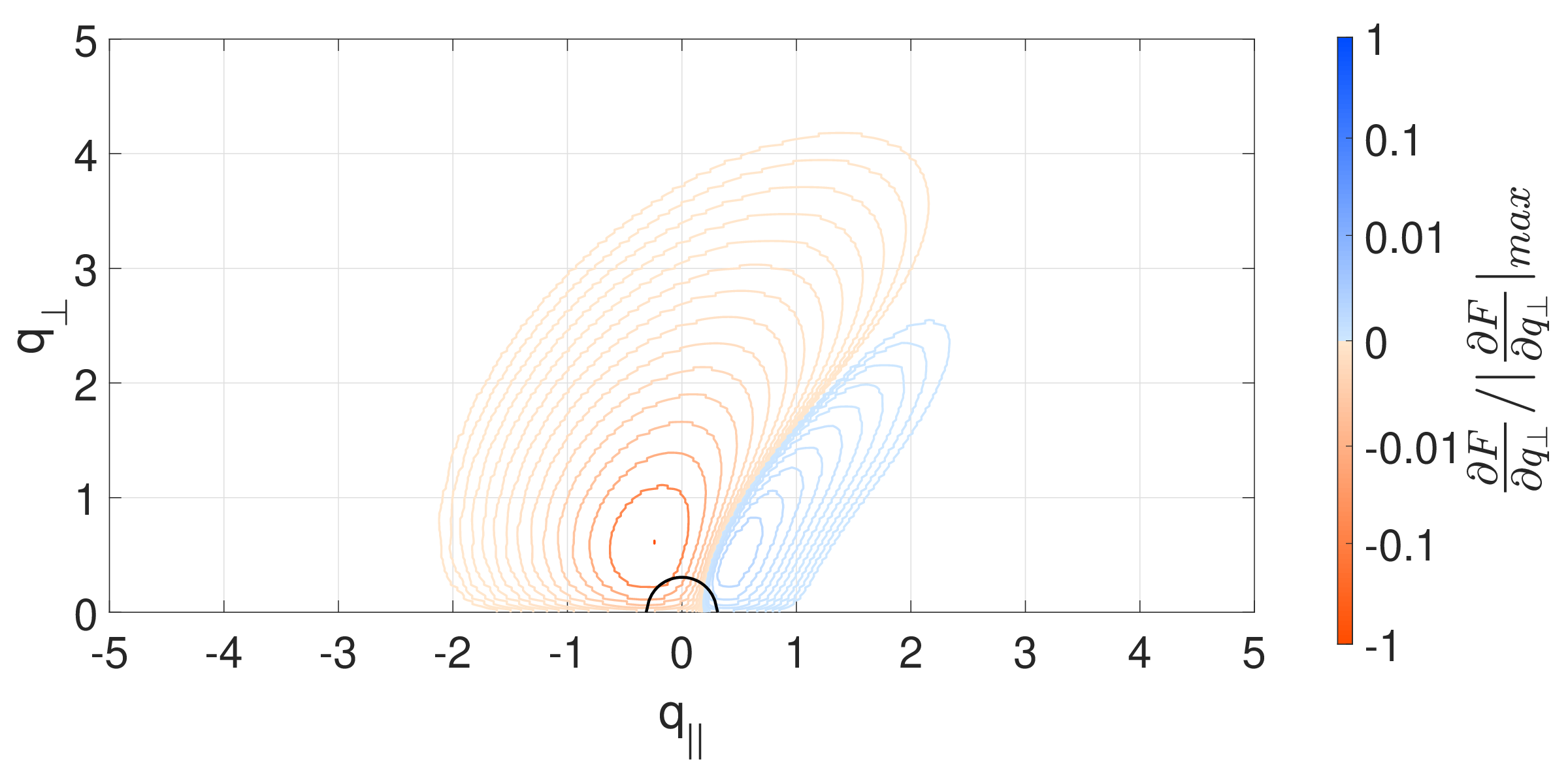}
	\caption{Contour plot showing $\frac{\partial F}{\partial q_\perp}$ for $\eta = 1$ and $\theta = 0.1$. The plot is normalised to $|\frac{\partial F}{\partial q_\perp}|_{max}$ for presentation purposes. Positive regions of the derivative are shown in blue, with negative regions in red. The black line shows the resonance ellipse for the slow X-mode with $\omega_r=0.95\Omega$ and $\phi = \pi/2$.  This ellipse produces a positive growth rate at lower temperatures but overlaps with too much of the $\frac{\partial F}{\partial q_\perp}<0$ area at higher temperatures such as this one. }
	\label{fig:res_plotb}
\end{figure}

 Sample resonance ellipses are shown on contour plots of $\partial F/\partial q_\perp $ for $\theta = 1/300$ in Fig. \ref{fig:res_plota} and $\theta =0.1$ in Fig. \ref{fig:res_plotb}, showing how the regions of negative gradient in the perpendicular direction cannot be avoided at higher temperatures due to their increased width. Although the growth rate calculation does not solely depend on $\partial F/\partial p_\perp$ (as shown in Eq. \ref{eq:lambda} and \ref{eq:Tn}), it is a good proxy for visualisation purposes.

\bsp	
\label{lastpage}
\end{document}